\documentclass[suppldata]{interact}
\usepackage[utf8]{inputenc}
\usepackage{xcolor}
\usepackage{ulem}
\newcommand{\stkout}[1]{\ifmmode\text{\sout{\ensuremath{#1}}}\else\sout{#1}\fi}
\usepackage{amsmath}
\usepackage{multirow}
\usepackage{array}
\usepackage{makecell}
\renewcommand\tablefont{\fontsize{10}{11}\selectfont}
\usepackage{graphicx}
\usepackage{siunitx}
\usepackage[export]{adjustbox}
\usepackage{xcolor}
\usepackage[numbers,sort&compress,merge]{natbib}
\bibpunct[, ]{[}{]}{,}{n}{,}{,}

\usepackage{caption}
\title{Embedded equation-of-motion coupled-cluster theory for electronic excitation, ionization, 
electron attachment, and electronic resonances}
\author{Valentina Parravicini and Thomas-C. Jagau \\[0.2cm] 
{\small Department of Chemistry, KU Leuven, Celestijnenlaan 200F, B-3001 Leuven, Belgium}}

\begin{document}

\maketitle

\begin{abstract}
The projection-based quantum embedding method is applied to a comprehensive set of electronic states. 
We embed different variants of equation-of-motion coupled-cluster singles and doubles (EOM-CCSD) 
theory in density functional theory and investigate electronically excited states of valence, Rydberg, and 
charge-transfer character, valence- and core-ionized states, as well as bound and temporary radical anions. 
The latter states, which are unstable towards electron loss, are treated by means of a complex-absorbing 
potential. Besides transition energies, we also present Dyson orbitals and natural transition orbitals for 
embedded EOM-CCSD. We benchmark the performance of the embedded EOM-CCSD methods against 
full EOM-CCSD using small organic molecules microsolvated by a varying number of water molecules as 
test cases. 

Our results illustrate that embedded EOM-CCSD describes ionization and valence excitation very well 
and that these transitions are quite insensitive towards technical details of the embedding procedure. 
On the contrary, more care is required when dealing with Rydberg excitations or electron attachment. 
For the latter type of transition in particular, the use of long-range corrected density functionals is 
mandatory and truncation of the virtual orbital space --which is indispensable for the application of 
projection-based embedding to large systems-- proves to be difficult. 
\end{abstract}


\section{Introduction} \label{sec:intro}
Electronic excitation, ionization, and electron attachment in complex environments are relevant to 
many frontier areas of chemical research including, among others, biochemistry \cite{Ali2015,Fab2017,
Vac2018}, utilization of solar energy \cite{Hes2018,Cas2018}, electrochemistry \cite{Mar2014}, and 
plasmonic chemistry \cite{Aslam2018}. The combination of large system sizes with small energy 
differences and subtle interactions poses a formidable challenge for electronic-structure theory and 
has provided the driving force for many methodological developments in recent 
years \cite{Kry2017,Lis2018,Gho2018,Lee2019,Izsak2020,Jones2020}. 

One potential approach to modeling complex systems, which motivates this work, is to start from a 
highly accurate \textit{ab initio} wave function method and to introduce approximations that lower the 
computational cost and at same time preserve the inherent accuracy of the method to the best possible 
degree. Among the different \textit{ab initio} approaches, the equation-of-motion coupled-cluster 
(EOM-CC) hierarchy of methods \cite{Emr1981,Sek1984,Sta1993,Nooi1993,Sta1994,Nooi1996,
Kry2008,Sne2012,ccbook} provides a useful framework for a unified treatment of electronic excitation, 
ionization, and electron attachment. EOM-CC theory is closely related to CC linear-response theory 
\cite{Mon1977,Muk1979,Dal1983,Koch1990} and holds several formal advantages: EOM-CC wave 
functions for different states are biorthogonal so that transition properties can be defined and evaluated 
in a straightforward manner. The same holds for quantities such as Dyson orbitals \cite{Lin1973,Ced1977,
Oan2007,Jag2016,Vid2020,Ort2020,Kry2020} and natural transition orbitals (NTOs) \cite{Luz1976,Luz1979,
Mar2003,Pla2014,Mew2018,Kry2020} that are useful for analyzing many-electron wave functions 
and also play a role for modeling photoionization spectra. Also, EOM-CC wave functions are spin 
eigenstates if a closed-shell state is used as reference and all EOM-CC transition energies are 
size-intensive. By including higher excitations in the wave-function ansatz, EOM-CC results can 
be systematically improved towards the exact solution \cite{Kal2004}. Through combination with 
complex variable techniques, that is, complex scaling \cite{Agu1971,Bal1971}, complex basis 
functions \cite{McC1978}, and complex absorbing potentials (CAPs) \cite{Riss1993}, EOM-CC 
theory has been extended to electronic resonances embedded in the continuum \cite{Gh2012,
Bra2013,Jag2014,Zuev2014,Whi2017,Jag2017}.

The computational cost of EOM-CC methods scales steeply with system size; already within the 
singles and doubles (EOM-CCSD) approximation as $\mathcal{O}(o^2 v^4)$ with $o$ as the number 
of occupied orbitals and $v$ the number of virtual orbitals. Since the accurate prediction of transition 
energies and properties requires fairly large basis sets, especially when dealing with Rydberg excited 
states or electronic resonances,\cite{Kau1989} the application of EOM-CCSD is restricted to relatively 
small systems. Over the years, many different strategies have been put forward to reduce the computational 
cost of the method; an excellent recent overview is provided by Ref. \citenum{Izsak2020}. One possible 
strategy to extend EOM-CC theory to larger systems is given by quantum embedding \cite{Wes1993,
Gov1998,Kni2012,Man2012,Lib2014,Wes2015,Lee2019,Jones2020}. Here, only a small region of a 
large system is described at the EOM-CC level of theory, whereas the remainder is treated in a more 
cost-efficient way. This is typically done by means of density functional theory (DFT), but the embedding 
of a higher-level CC model in a lower-level one is also possible.\cite{Koch2014,Cou2018}

Among the different wave-function in DFT embedding methods \cite{Gov1998,Lib2014,Lee2019,
Jones2020}, projection-based techniques \cite{Man2012,Kha2012,Good2014,Heg2016,Chu2017,
Cul2017} have recently gained popularity. A recent account of projection-based embedding and an 
overview of the wide range of systems to which it has been applied is provided by Ref. \citenum{Lee2019}. 
In this method, the influence of the environment on the high-level fragment is taken into account by 
means of an embedding potential that is determined from the density of the full system. A particular 
advantage of the method is that orbitals assigned to different fragments are orthonormal to each other. 
This ensures that the energy of the full system is recovered exactly if both fragments are described 
at the same level of theory and also has a positive effect on the more common case where a high-level 
wave-function method such as EOM-CC is embedded in DFT. A further advantage of projection-based 
embedding is that no modifications are necessary to subsequent CC and EOM-CC calculations as 
long as orbital relaxation is not considered; they will just be performed with fewer occupied orbitals. 
However, for the method to be efficient it is critical to prune the virtual orbital space or, alternatively, 
the basis set as well and several ideas have been put forward in this context \cite{Bar2013,Ben2015,
Heg2018,Ben2019,Cla2019}. Further methodological developments in the context of projection-based 
embedding include the combination with classical mechanics in a QM/MM framework \cite{Ben2016,
Rana2019}, the application to periodic systems \cite{Chu2018}, and analytic nuclear gradients \cite{Lee2019g}. 

Progress in the application of wave-function in DFT embedding to excited states and electron-detached 
and attached states has been slower \cite{Izsak2020,Hoe2012,Hoe2013,Dad2013,Dad2014,Pra2016,
Ben2017,Ric2018,Wen2020}. One central question is whether the use of the same embedding potential 
for different electronic states is appropriate, that is, whether a state-specific or a state-universal approach 
is superior. Recent investigations using projection-based embedded EOM-CCSD \cite{Ben2017} suggest 
that a state-universal approach is valid at least for valence excitations in microsolvated small molecules. 
Similar applications to larger molecules \cite{Wen2020} using the absolutely localized variant of 
projection-based embedding \cite{Chu2017} point in the same direction. The combination with 
complete-active-space self-consistent-field methods is worth noting in this context as well \cite{Lima2017}. 
Ionization energies have also been computed with projection-based embedding \cite{Bar2015} but 
only in a state-specific manner through separate CC calculations for the neutral and ionized states 
and not by means of EOM-CC. There are, however, corresponding developments for electron-detached 
and attached states in the context of frozen-density embedding of EOM-CCSD \cite{Bou2018} and the 
second-order algebraic diagrammatic construction scheme \cite{Liu2020}.


In this work, we extend projection-based embedding to the EOM-CCSD variants for ionization and 
electron attachment. Through the combination with CAPs we are also able to apply the method to electronic 
resonances. Furthermore, we introduce NTOs and Dyson orbitals for embedded EOM-CCSD. A second 
objective of our article is the comprehensive assessment of embedded EOM-CCSD for different types 
of states. We report transition energies for valence and Rydberg excitations, charge transfer states, 
valence and core ionizations, and electron attachment resulting in both bound and temporary anions. 
These energies are compared to results from regular EOM-CCSD as well as time-dependent (TD)-DFT 
and $\Delta$DFT calculations.  

The article is organized as follows: Section \ref{sec:comp} gives a brief account of the theory underlying 
our work and the technical details of our computations. Section \ref{sec:results1} presents our results 
for different types of bound states, while Section \ref{sec:results2} deals with electronic resonances. 
Our concluding remarks are given in Section \ref{sec:conc}.


\section{Computational methods} \label{sec:comp}
We combine in this work quantum embedding based on a level-shift projector \cite{Man2012,Lee2019} 
with various flavours of EOM-XX-CC theory and calculate excitation energies (XX=EE), ionization 
potentials (XX=IP), and electron attachment energies (XX=EA). General overviews of CC theory 
and the different flavors of EOM-CC theory are, for example, provided in Refs. \citenum{ccbook,Kry2008,
Sne2012}. Computational strategies towards larger molecules in the context of EOM-CC have been 
reviewed recently in Ref. \citenum{Izsak2020}. More details about the treatment of electronic resonances 
in EOM-CC theory can be found in Ref. \citenum{Jag2017}. In the following, we give a brief account of 
some aspects that are specific to our calculations. 

\subsection{Projection-based embedding} \label{sec:pemb}
\begin{figure} \centering
\includegraphics[scale=0.72,center]{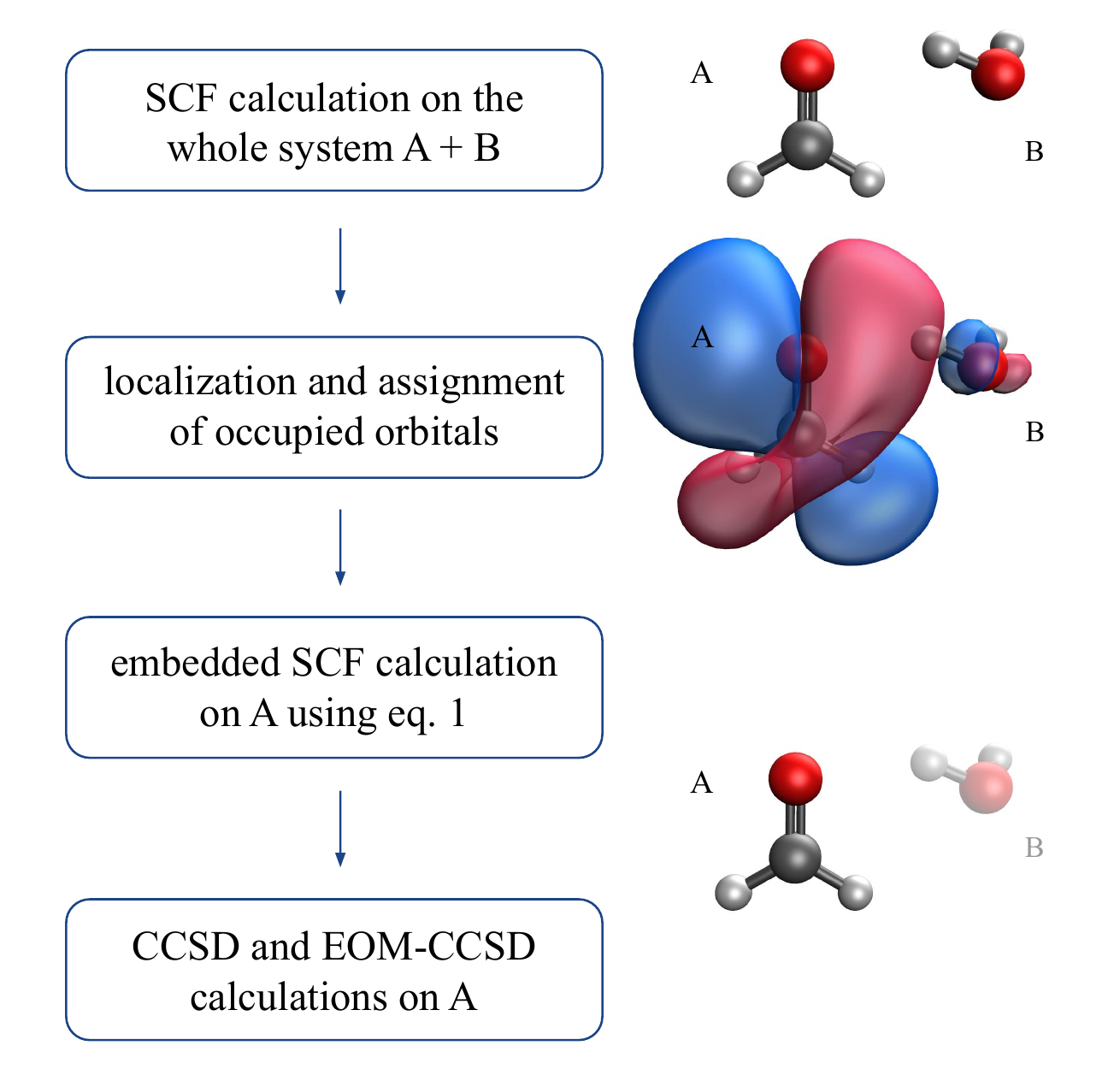}
\caption{Flow of action in an embedded EOM-CCSD calculation.}
\label{fig:pemb}
\end{figure}

We partition a chemical system into a subsystem \textbf{A}, whose energy or properties we are primarily 
interested in and which is treated using EOM-CC theory, and its environment \textbf{B}, which is treated 
with DFT. The flow of a calculation is illustrated in Figure \ref{fig:pemb}. First, we perform a standard DFT 
calculation for the whole system \textbf{A+B}. The resulting density matrix is split into two pieces 
$\boldsymbol{\gamma}^\text{A}$ and $\boldsymbol{\gamma}^\text{B}$ after localization of the occupied 
orbitals. In this work, we do this by means of a Mulliken population analysis \cite{Mul1955} and Pipek-Mezey 
localization \cite{Pip1989}. 

Subsequently, the density matrix for the embedded fragment \textbf{A} is re-computed in a second 
self-consistent field calculation \cite{Man2012}. This is done using the modified Fock matrix
\begin{align} \label{eq:emb}
\widetilde{\mathbf{F}}^\text{A-in-B}[\boldsymbol{\gamma}^\text{A}] &= 
(\mathbf{I}-\mathbf{S}\boldsymbol{\gamma}^B) \, 
\mathbf{F}^\text{A-in-B}[\boldsymbol{\gamma}^\text{A}] \, (\mathbf{I}-\boldsymbol{\gamma}^B\mathbf{S})~, \\
\mathbf{F}^\text{A-in-B}[\boldsymbol{\gamma}^\text{A}] &= \mathbf{h} + 
\mathbf{g}[\boldsymbol{\gamma}^\text{A}] + \mathbf{v}_\text{emb}^\text{A-in-B} \nonumber
\end{align}

by projecting out the occupied orbitals assigned to fragment \textbf{B}, and thereby 
enforcing orthonormality among all occupied orbitals. Here, \textbf{h} denotes the one-electron Hamiltonian, 
\textbf{g} the mean-field two-electron potential, and \textbf{S} the overlap matrix. The embedding potential 
is given as $\mathbf{v}_\text{emb}^\text{A-in-B} = \textbf{g}[\boldsymbol{\gamma}^\text{A+B}] - 
\textbf{g}[\boldsymbol{\gamma}^\text{A}]$. This approach is equivalent to the original 
formulation \cite{Man2012,PEMB} where the projector was chosen as $\mu\, \mathbf{S} \boldsymbol{\gamma}^\text{B} 
\mathbf{S}$ with $\mu$ as a large constant. The converged density matrix $\boldsymbol{\gamma}^\text{A}$ 
forms the basis for the subsequent EOM-CCSD treatment of subsystem \textbf{A}. No modifications of the 
CCSD and EOM-CCSD equations are necessary to this end.

To truncate the virtual orbital space, we employ concentric localization \cite{Cla2019}: In this approach, 
the full virtual orbital space is projected on fragment \textbf{A} and an initial set of virtual orbitals is 
obtained from a singular value decomposition of the overlap between the full and the projected virtual 
space. The full virtual space can then be reconstructed in an iterative fashion by analyzing the coupling 
of the orbital spaces through the Fock operator giving tunable accuracy. In this work, the virtual space 
consists of the initial and one further set of orbitals unless indicated otherwise. 

\subsection{Dyson orbitals and natural transition orbitals} \label{sec:orbs}
Similar to other quantum chemistry methods, it is also in the framework of projection-based embedding 
important to characterize changes in the wave function upon excitation, ionization, or electron attachment. 
Two quantities that are useful in this context are Dyson orbitals \cite{Lin1973,Ced1977,Oan2007,Jag2016,
Vid2020,Ort2020,Kry2020} and natural transition orbitals (NTOs) \cite{Luz1976,Luz1979,Mar2003,Pla2014,
Mew2018,Kry2020}. Within projection-based embedded EOM-CC, we can define and evaluate these 
quantities in analogy to regular EOM-CC. Dyson orbitals $\phi_d$ are given as 
\begin{equation} \label{eq:dyson}
\phi_d (1) = \sqrt{n} \; \int \!\! \mathrm{d}1 \, \mathrm{d}2 \dots \mathrm{d}n \;\; \Psi_n^*(1, 2, \dots, n) 
\;\; \Psi_{n-1} (2, \dots, n)
\end{equation}
and can be viewed as generalized transition density matrices between a pair of states with $n$ 
and $n-1$ electrons, respectively. EOM-CC Dyson orbitals thus characterize ionization or electron 
attachment based on the correlated many-electron wave functions taking part in the transition. Notably, 
Eq. \eqref{eq:dyson} is not limited to the ground state but can be applied to ionization or attachment 
involving excited states as well \cite{Oan2007}. Because of the asymmetry of the similarity transformed 
Hamiltonian in EOM-CC theory, pairs of left and right Dyson orbitals exist but they usually differ only 
marginally. 

NTOs arise from a singular value decomposition of the transition density matrix $\boldsymbol{\gamma}^{IF}$ 
for a pair of states $\Psi_I$ and $\Psi_F$, 
\begin{align} \label{eq:nto}
\boldsymbol{\gamma}^{IF}(1,1') &= n \int \!\! \mathrm{d}1 \, \mathrm{d}2 \dots \Psi_I^* (1, 2, \dots, n) \; 
\Psi_F (1', 2, \dots, n) \\
&= \sum_K \sigma_K \psi^p_K (1) \psi^h_K(1') ~, \nonumber \\
\psi_K^p (1) &= \sum_q \, V_{qK} \, \phi_q (1) ~, \quad \psi_K^h (1) = 
\sum_q \, U_{qK} \, \phi_q (1)~. \nonumber 
\end{align}
Here, $\psi_K^p$ and $\psi_K^h$ denote particle and hole NTOs, respectively, while $\phi_q$ is a 
generic molecular orbital. The singular values $\sigma_K$ provide a measure of the contribution of 
a particular NTO pair to the overall transition and usually only very few of them are large. In EOM-CC 
theory, pairs of NTOs exist because $\boldsymbol{\gamma}^{IF} \neq \boldsymbol{\gamma}^{FI}$ 
but the differences are again marginal. 

\subsection{Electronic resonances} \label{sec:res}
Electronic resonances are states that are metastable towards electron loss \cite{Jag2017,nhqmbook}. 
Such states cannot be tackled with conventional quantum chemistry methods because their wave 
functions are not square integrable in Hermitian quantum mechanics. Within non-Hermitian quantum 
mechanics,\cite{nhqmbook} however, the energy of a resonance can be defined as
\begin{equation} \label{eq:siegert}
E = E_R - i\, \Gamma/2~, 
\end{equation}
where the real part is the position of the resonance and the imaginary part the half-width, which is 
related to the state's lifetime, $\tau=1/\Gamma$. Among the different non-Hermitian approaches that 
were introduced for the treatment of electronic resonances \cite{Jag2017,nhqmbook}, we use here 
the complex absorbing potential (CAP) approach \cite{Riss1993}. Here, the Hamiltonian is augmented 
by an artificial potential $W$ that absorbs the diverging tail of the resonance wave function. The energy 
eigenvalues of the CAP Hamiltonian $H(\eta) = H - i \, \eta \, W$ will depend on the shape of the potential 
$W$ and on the CAP strength $\eta$. Its optimal value $\eta_{opt}$ is found by computing energies for 
a range of $\eta$ values and minimizing the quantity $|\eta \, dE/d\eta|$ \cite{Riss1993}. In this work, 
shifted quadratic CAPs are used and the CAP-free region is determined either based on Voronoi cells  
\cite{Som2015} or takes the form of a box.

Different variants of CAP-EOM-CC methods can be distinguished depending on whether the CAP 
is included already in the Hartree-Fock (HF) calculation, or at the CCSD level, or only at the EOM-CCSD 
level.\cite{Jag2017} In this work, the third approach is employed, that is, the CAP is added only at the 
EOM-CCSD level. In addition, the CAP is projected on the virtual orbital space \cite{San2001}. Although 
this approach is less consistent than the first one from a formal standpoint \cite{Jag2017}, it offers an 
advantage in the context of quantum embedding: since the HF reference wave function is real-valued, 
the embedding procedure need not be adapted to non-Hermitian quantum chemistry. Moreover, the 
computational cost is considerably lower because only a single CAP-free CCSD calculation (scaling 
as $\mathcal{O}(o^2\, v^4)$) is needed for the reference state, whereas the EOM-CCSD part, which 
scales as $\mathcal{O}(o\, v^4)$ for electron attachment, needs to be repeated for different values 
of $\eta$. It can also be argued that the CAP represents nothing but a perturbation for the bound 
reference state that should be avoided. 

\subsection{Technical details} \label{sec:compdet}
In all computations except those for charge transfer states, the environment \textbf{B} consists of a 
varying number (1--5) of water molecules, whereas we take different small molecules as fragment 
\textbf{A}: formaldehyde (CH$_2$O), methanol (CH$_3$OH), ethylene (C$_2$H$_4$), methylene 
(CH$_2$), fluoromethylene (HCF), silylene (SiH$_2$), dinitrogen (N$_2$), and carbon monoxide 
(CO). These systems are small enough to be treated as a whole by EOM-CCSD and the comparison 
between full and embedded EOM-CCSD is the centerpiece of our work. Furthermore, we report 
transition energies obtained with computationally cheaper alternatives: Ionization and electron-attachment 
energies are calculated with $\Delta$DFT using the maximum overlap method (MOM) \cite{Gil2008} 
and for excited states TD-DFT calculations are carried out within the Tamm-Dancoff approximation 
(TDA/TD-DFT)\cite{Run1984,Hir1999}. As concerns charge transfer, we investigate two systems: 
HCl (\textbf{A}) embedded in 5 H$_2$O (\textbf{B}), where charge transfer takes place in the HCl 
molecule, and the donor-acceptor complex NH$_3\cdots$F$_2$ where we take the donating NH$_3$ 
molecule as fragment \textbf{A} and the accepting F$_2$ molecule as fragment \textbf{B}.

The structures of the systems consisting of two molecules, that is, \textbf{A} + H$_2$O and the CO$_2$ 
and N$_2$ dimers, are optimized with RI-MP2/cc-pVTZ unless stated otherwise. Clusters composed 
of four and six molecules, that is, fragment \textbf{A} + 3 or 5 H$_2$O, are generated starting from the 
optimized structure of \textbf{A}. All geometries can be found in the Supporting Information (SI). The 
following density functionals are used as low-level method in the embedded EOM-CCSD calculations 
and also for standalone $\Delta$DFT and TD-DFT calculations: PBE \cite{Per1996} and BLYP \cite{
Bec1988, Lee1988} (generalized gradient approximation (GGA) functionals), PBE0 \cite{Ada1999} 
(hybrid GGA functional), as well as CAM-B3LYP \cite{Yan2004}, $\omega$B97X-D \cite{Chai2008}, 
and LC-$\omega$PBE08 \cite{Wei2009} (range-separated hybrid functionals). The subsequent CCSD 
and EOM-CCSD calculations are always based on a HF reference wave function for the embedded 
fragment \textbf{A} and the core orbitals of \textbf{A} are always included in the correlation treatment 
except for core ionization energies. Here, the frozen-core/core-valence-separation (fc-CVS) scheme 
\cite{Ced1980,Cor2015,Vid2019} is employed for the clusters comprising more than one water molecule 
because convergence of the EOM-CCSD equations is not achieved otherwise.

The correlation-consistent basis set aug-cc-pVTZ \cite{Dunn1992} is employed for most systems. 
For temporary anions and Rydberg excited states, respectively, this basis is augmented with 3 or 
2 additional $s$ and $p$ shells of diffuse functions. These extra shells are generated in an 
even-tempered manner, i.e., by recursively dividing the exponent of the most diffuse $s$ and $p$ 
shell by a factor of 2. All calculations were performed with a slightly modified version of the Q-Chem 
electronic structure package \cite{qchem,qchem5} using the implementation of projection-based 
embedding included in the 5.3 release of the program \cite{PEMB}. Results of embedded EOM-EE-CCSD 
calculations were verified against results presented in Ref. \citenum{Ben2017}.


\section{Results for bound states} \label{sec:results1}
\subsection{Performance of embedded EOM-CCSD at different distances between the fragments}

\begin{figure} \centering
\includegraphics[width=1.2\textwidth,center]{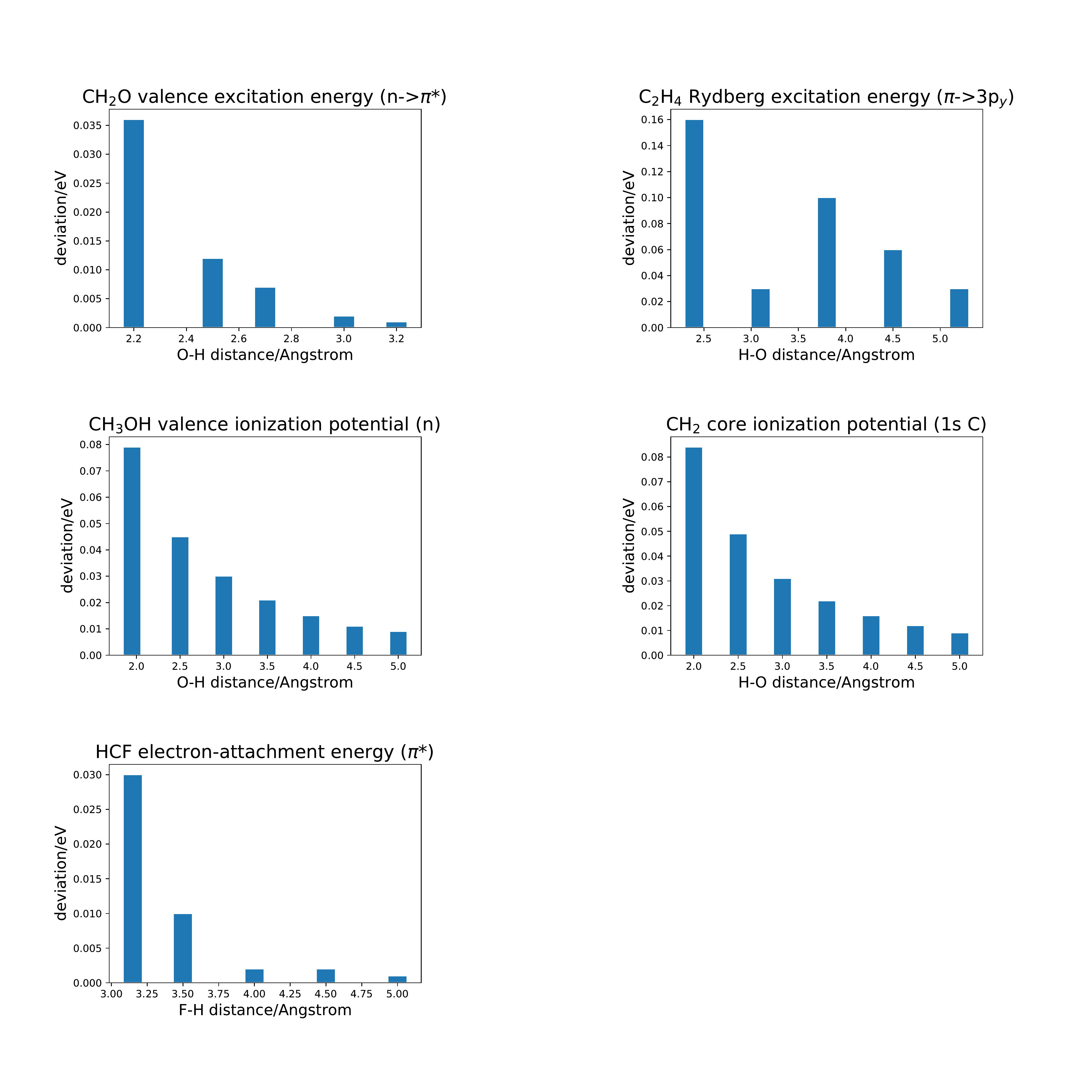}
\caption{Absolute value of the deviation between transition energies computed with EOM-CCSD 
embedded in PBE0 and with full EOM-CCSD. All systems are composed of a molecule in which 
the excitation takes place and a single water molecule that acts as environment. The deviation is 
shown as a function of the distance between the two fragments.}
\label{fig:dev-dist}
\end{figure}

Figure \ref{fig:dev-dist} presents deviations of embedded EOM-CCSD from full EOM-CCSD as 
a function of the intermolecular distance between the fragments. Representative examples of a 
valence excitation, a Rydberg excitation, a valence ionization, a core ionization, and an electron 
attachment are shown; the lower-level fragment always consists of a single H$_2$O molecule. 
In all these calculations, the full virtual orbital space is taken into account in the embedded 
EOM-CCSD calculation. 

Figure \ref{fig:dev-dist} illustrates that the deviations between embedded and full EOM-CCSD 
decrease with the intermolecular distance and stay below 0.1 eV even at the shortest distance. 
This holds true for every type of transition that we examined with the notable exception of Rydberg 
excitations. In this case, not only is the trend not observed anymore, but also the absolute deviations 
are larger and the performance depends on the orbital to which the electron is excited. This might 
be tied to the spatial extension of Rydberg states, which are perturbed more strongly by the presence 
of the water molecule also at a larger distance, and are more sensitive to the relative orientation. 
We note that Bennie \textit{et al.} observed as well that the deviation between embedded and full 
EOM-CCSD decreases with growing intermolecular distance for valence excited states \cite{Ben2017}. 

The same trend is observed in larger clusters where the lower-level fragment comprises 5 H$_2$O 
molecules. In Tables \ref{5-ee} and \ref{5-ip}, deviations between embedded and full EOM-CCSD 
are shown for a valence excitation of CH$_2$ and a valence ionization of C$_2$H$_4$, respectively. 
Two configurations \textit{near} and \textit{far} are compared: In the first cluster, the distances 
between all water molecules and the excited or ionized molecule lie around 2--3 \si{\angstrom}, in 
the other cluster they are of the order of 5 \si{\angstrom}. The errors differ heavily between the 
two clusters: For the configuration \textit{far}, they are smaller by a factor of up to 100 as compared 
to the configuration \textit{near}. We note that the manual inclusion of additional occupied orbitals 
in the high-level fragment can improve the performance of embedded EOM-CCSD at short 
distances \cite{Ben2017}, this is however not explored further here. 


\subsection{Influence of the size of the environment region} 
Tables \ref{5-ee}--\ref{5-ea} report deviations between embedded and full EOM-CCSD for clusters 
in which the environment consists of three or five water molecules. The results are grouped by type 
of state, negative deviations correspond to a transition energy lower than that of full EOM-CCSD. 
For these clusters, the distances between the water molecules and the embedded fragment vary 
among the configurations, but they usually lie in the lower half of the distances presented in Figure 
\ref{fig:dev-dist} at which correlation effects are more important and the performance of the embedded 
method is less good.

As we can see from the comparison of error values reported in Figure \ref{fig:dev-dist} and Tables 
\ref{5-ee}--\ref{5-ea}, for the same type of transition and the same molecule, the error due to embedding 
generally increases with the amount of water molecules in the environment. The performance differs 
considerably among the individual molecules and also depends on the density functional used to 
treat the environment, but the deviations between embedded and full EOM-CCSD for the clusters 
with three or five water molecules can exceed those for the two-molecule clusters by a factor of 10 
and more. 

On the other hand, the configuration of the larger clusters has a relatively small effect on the deviation 
of the embedded energies, provided that the distances between the water molecules and the high-level 
fragment lie in the same range. Electron attachment (see Table \ref{5-ea}) shows the highest sensitivity: 
Configurations 1, 2 and 4 of HCF, where embedded EOM-EA-CCSD performs particularly poorly, have 
in common a single water molecule positioned such that the electron density of the oxygen lone pair 
faces the halocarbene.


\subsection{Performance for different types of target states}
In this section, we analyze the performance of projection-based embedded EOM-CCSD with respect 
to the type of transition. This is again done based on the data presented in Tables \ref{5-ee}--\ref{5-ea}.

\begin{table}\setlength{\tabcolsep}{2.5pt}
\tbl{Valence excitation energies calculated with full EOM-EE-CCSD, embedded EOM-EE-CCSD, and 
TD-DFT using the aug-cc-pVTZ basis set. Embedded EOM-EE-CCSD and TD-DFT results (last three 
columns) are reported as deviations from full EOM-EE-CCSD. All values in eV.} 
{\begin{tabular}{lcccccc} \toprule
System & Transition & Configuration & \makecell{full \\ EOM-CCSD} &  \makecell{EOM-CCSD 
\\ in PBE} & \makecell{EOM-CCSD \\ in CAM-B3LYP} & \makecell{TDA/ \\ TD-CAM-B3LYP} \\  
\midrule
CH$_2$O + 5H$_2$O  & n$\rightarrow$ $\pi$* & 1  & 4.08 & 0.04    & 0.05 & -0.06 \\
&& 2                     & 4.02 & 0.001  & 0.01 & -0.05 \\
&& 3                     & 3.92 & -0.008 & 0.004 & -0.06 \\
&& 4                     & 4.00 & 0.02   & 0.02 & -0.05 \\
&& 5                     & 3.94 & 0.008   & 0.01 & -0.06 \\[0.3cm]
CH$_3$OH  + 5H$_2$O  & n$\rightarrow$$\sigma$*& & 7.04 & -0.54 &  -0.07 & -0.64 \\[0.3cm]
HCF + 5H$_2$O & $\pi$*$\rightarrow$$\pi$* &  & 3.14 & 0.36   & 0.20 & 0.40 \\[0.3cm]
CH$_2$  + 5H$_2$O & $\sigma$ $\rightarrow$ $\pi^*$ & \textit{near} & 1.70  & 0.29  & 0.27 & -0.07 \\
       & & \textit{far} & 1.66  & 0.002   &  0.004 & -0.07 \\
\bottomrule
\end{tabular}}
\label{5-ee}
\end{table}

\textbf{Valence excitations.} As we can see from Table \ref{5-ee}, excitation energies obtained with 
EOM-EE-CCSD embedded in PBE and CAM-B3LYP vary appreciably among the test cases. The 
two density functionals produce similar results in general, but the excitation energy can differ by up 
to 0.47 eV (CH$_3$OH + 5H$_2$O). Also, the errors of embedded EOM-EE-CCSD are comparable 
to the ones obtained with TD-CAM-B3LYP in most cases. We add that Bennie \textit{et al.} presented 
a comparison to different density functionals as well and found that time-dependent PBE, PBE0, 
M06-2X, and HF perform significantly worse than embedded EOM-EE-CCSD in predicting the 
excitation energy of microsolvated formaldehyde and acroleine \cite{Ben2017}. An exception is 
found among our results for CH$_3$OH, where EOM-CCSD embedded in CAM-B3LYP performs 
very well, whereas TD-CAM-B3LYP struggles to reproduce the full EOM-CCSD energy. This might 
be explained by the partial Rydberg character (3$s$) of this excited state \cite{Lan2020} whose 
proper representation is difficult with TD-DFT. 


\begin{table}[h] \setlength{\tabcolsep}{3pt}
\tbl{Rydberg excitation energies calculated with full EOM-EE-CCSD, embedded EOM-EE-CCSD, 
and TD-DFT using the aug-cc-pVTZ+2s2p basis set. Embedded EOM-EE-CCSD 
and TD-DFT results (last three columns) are reported as deviations from full EOM-EE-CCSD. All 
values in eV.}
{\begin{tabular}{lcccccc} \toprule
System  & Transition\textsuperscript{a}&  \makecell{full \\ EOM-CCSD} &\makecell{EOM-CCSD \\ 
in PBE} & \makecell{EOM-CCSD \\ in CAM-B3LYP} &  \makecell{TDA/ \\ TD-CAM-B3LYP} \\  
 \midrule
 CH$_2$O + 3H$_2$O & n$\rightarrow$3s & 7.69 & -0.17 & 0.05 & -0.33 \\  
                    & n$\rightarrow$3p$_z$ & 8.42 & -0.19 & 0.01 & -0.27 \\
                    & n$\rightarrow$3p$_y$ & 8.70 & -0.36 & 0.03 & -0.39 \\
                    & n$\rightarrow$3p$_x$  & 8.82 & -0.13 & 0.05 & -0.43 \\[0.3cm]
CH$_3$OH + 3H$_2$O & n$\rightarrow$3p$\sigma$& 8.37 &-0.27&-0.01&-0.60 \\
                     & n$\rightarrow$3p$\pi$& 8.75  &-0.48&-0.03&-0.68 \\
                     & n$\rightarrow$3p$\sigma^{\prime}$& 8.85 &-0.52&-0.08& -0.81 \\[0.3cm]
C$_2$H$_4$ + 3H$_2$O & $\pi\rightarrow$3s &7.06 &-0.21& 0.09 &-0.65 \\
                     & $\pi\rightarrow$3p$_y$ & 7.74 &-0.15& 0.04 & -0.94 \\
                     & $\pi\rightarrow$3p$_z$ & 7.76 &-0.08& 0.11 &-0.74 \\
                     & $\pi\rightarrow$3p$_x$ & 7.92 &-0.02& 0.12 &-0.80 \\
\bottomrule
\end{tabular}}
\tabnote{\textsuperscript{a}Assignment follows Ref. \citenum{Lis2002} for CH$_2$O, Ref. 
\citenum{Lan2020} for CH$_3$OH and Ref. \citenum{Fel2014} for C$_2$H$_4$.}
\label{5-ry}
\end{table}
 
\textbf{Rydberg excitations.} Due to the incorrect asymptotic behaviour of local exchange-correlation 
functionals, lower-rung TD-DFT methods usually offer only a poor description of Rydberg excited 
states \cite{Dre2003}. The performance improves when range-separated functionals are employed 
\cite{Car2010}. However, in Table \ref{5-ry} we can see than TD-CAM-B3LYP still deviates by 0.3--1.0 
eV from full EOM-EE-CCSD. This is in contrast to EOM-EE-CCSD embedded in CAM-B3LYP, where 
deviations do not exceed 0.12 eV. In contrast to valence excitations, embedded EOM-CCSD thus 
represents a clear improvement over TD-CAM-B3LYP for Rydberg excitations. As observed for 
valence excited states as well, the choice of the low-level method has an impact on the results of 
embedded EOM-EE-CCSD: CAM-B3LYP produces significantly smaller deviations than PBE. This 
can be ascribed to the long-range HF exchange included in the former density functional and mimics 
the performance of the corresponding TD-DFT methods.  

 
\begin{table}
\tbl{Excitation energies of charge transfer states calculated with full and embedded 
EOM-EE-CCSD/aug-cc-pVTZ and TD-DFT. All values in eV.}
{\begin{tabular}{lcccccccc} \toprule
System & Transition & \makecell{full \\ EOM-\\CCSD} &  \makecell{EOM-\\CCSD 
\\ in PBE}  &  \makecell{EOM-\\CCSD \\ in BLYP} & \makecell{EOM-CCSD \\ in CAM-\\B3LYP} & 
\makecell{TDA/ \\ TD-\\PBE} & \makecell{TDA/ \\ TD-\\ BLYP} & \makecell{TDA/ \\ TD-CAM-\\ B3LYP} \\  
\midrule
HCl + 5H$_2$O & $n \rightarrow \sigma$* & 8.19 & 8.00 & 8.15 &8.18 & 6.92&6.80 & 7.80 \\ 
NH$_3$$\cdots$F$_2$  & $n \rightarrow \sigma$* & 4.49 & 3.12\textsuperscript{a} & 
5.27\textsuperscript{a} & 6.51\textsuperscript{a} & 0.31 & 0.28 & 4.87\\
\bottomrule
\end{tabular}}
\tabnote{\textsuperscript{a}Virtual orbital space is not truncated.}
\label{ct}
\end{table}

\textbf{Charge transfer states.} Our results for intra- and intermolecular charge transfer are reported 
in Table \ref{ct}. Embedded EOM-EE-CCSD describes the intramolecular process that occurs 
in HCl + 5H$_2$O with good accuracy; the excitation energy deviates only by 0.01 eV from full 
EOM-CCSD when CAM-B3LYP is used as low-level method. Although embedded EOM-EE-CCSD 
is more accurate than TD-DFT also with PBE and BLYP as low-level method, the dependence of 
the results on the density functional is not eliminated. PBE yields the largest error, while BLYP 
gives intermediate results that are slightly improved by error cancellation when the virtual orbital 
space is truncated. These results can be related to the fact that local density functionals do not 
provide a good description of charge transfer.\cite{Dre2003} 

As a test case for intermolecular transfer, we investigate the NH$_3 \cdots$F$_2$ complex that 
was used as a test case for the M06-HF density functional \cite{Zha2006}. The same geometry 
is used here. The complex is divided such that the donating fragment (NH$_3$) is treated at the 
EOM-CCSD level, whereas the accepting one (F$_2$) constitutes the environment and is treated 
at the DFT level. We do not truncate the virtual orbital space in the embedded EOM-CCSD 
calculations so that it remains identical to that employed in the full EOM-CCSD calculations. 
However, even though the virtual orbital space is the same, embedded EOM-EE-CCSD does 
not perform well, the errors with respect to full EOM-EE-CCSD amount to up to 2 eV. It is also 
remarkable that embedding in CAM-B3LYP produces the worst results, while the lower-rung 
functionals BLYP and PBE lead to better results. Within TD-DFT, however, CAM-B3LYP provides 
an accurate description, whereas the GGA functionals are completely incapable of describing the 
charge transfer. We conclude that embedded EOM-CCSD provides a reliable description of charge 
transfer only if both the donor and the acceptor are part of the high-level fragment. 


\begin{table}[h] \setlength{\tabcolsep}{3pt}
\tbl{Valence ionization energies calculated with full EOM-IP-CCSD, embedded EOM-IP-CCSD, and 
$\Delta$DFT using the aug-cc-pVTZ basis set. Embedded EOM-IP-CCSD and $\Delta$DFT results 
(last four columns) are reported as deviations from full EOM-IP-CCSD. All values in eV.}
{\begin{tabular}{lccccccc} \toprule
System & Transition &Configuration & \makecell{full \\EOM-CCSD} & \makecell{EOM-CCSD \\in PBE} & 
\makecell{EOM-CCSD \\in CAM-B3LYP} & \makecell{$\Delta$CAM-\\B3LYP} & $\Delta$PBE\\  
\midrule
CH$_2$O + 5H$_2$O & $\pi$*$\rightarrow$$\infty$ &1  & 11.20 & 0.22 & 0.23 
&  -0.24 & -0.15\\
&& 2                    & 11.19 & 0.10 & 0.12  &-0.15 & -0.22 \\
&& 3                    & 11.16 & 0.10 & 0.12  &-0.20 & -0.18 \\
&& 4                    & 11.70 & 0.09 & 0.14  &-0.45 & 0.66  \\
&& 5                    & 11.12 & 0.14 & 0.15  & -0.12 & -0.15 \\[0.3cm]
HCF + 5H$_2$O &n$\rightarrow$$\infty$&& 10.16 & 0.37  & 0.56 & -0.32 & --\textsuperscript{a}\\[0.3cm]
CH$_3$OH  + 5H$_2$O  & n$\rightarrow$$\infty$ && 10.47 & 0.29  & 0.30 & 0.17 & 3.77 \\[0.3cm]
C$_2$H$_4$  + 5H$_2$O & $\pi$$\rightarrow$$\infty$ & \textit{near} & 8.98 & 0.64 & 0.64 & 0.23 & -0.26 \\
                     & & \textit{far}  & 10.94 & 0.02 & 0.03 &  -0.10& -0.10 \\
\bottomrule
\end{tabular}}
\tabnote{\textsuperscript{a}PBE calculation of the ionized state does not converge.}
\label{5-ip}
\end{table}

\textbf{Valence ionizations.} As apparent from the comparison of Tables \ref{5-ee} and \ref{5-ip}, 
the deviations between embedded and full EOM-CCSD are somewhat larger for the IP variant 
than for the EE variant. Similar to valence excitations, the dependence on the density functional 
used for the description of the environment is weak but significant differences are obtained in 
some cases. Among the cases we considered, the largest difference of ca. 0.2 eV occurs for 
HCF+5 H$_2$O. Remarkably, embedding in the lower-rung functional PBE produces better 
results for this case than embedding in CAM-B3LYP. By contrast, $\Delta$CAM-B3LYP is clearly 
superior to $\Delta$PBE and performs similar to embedded EOM-IP-CCSD. This implies that 
embedded EOM-IP-CCSD does not always represent an improvement over $\Delta$CAM-B3LYP 
if one is interested only in the ionization energies. However, a distinct advantage of all EOM-CC 
methods is that the wave functions for the ground and target state are biorthogonal so that 
transition properties can be readily evaluated. This is not the case in $\Delta$DFT approaches 
where the evaluation of further quantities besides the transition energy is not straightforward.


\begin{table}[h] \setlength{\tabcolsep}{4pt}
\tbl{Core ionization energies calculated with full EOM-IP-CCSD, embedded EOM-IP-CCSD, and 
$\Delta$DFT using the aug-cc-pVTZ basis set. Embedded EOM-IP-CCSD and $\Delta$DFT results 
(last three columns) are reported as deviations from full EOM-IP-CCSD. Frozen core-CVS scheme 
is employed for full and embedded EOM-IP-CCSD calculations. All values in eV.}
{\begin{tabular}{lcccccc} \toprule
System & Transition & Configuration &\makecell{full \\ EOM-CCSD} &\makecell{EOM-CCSD \\ 
in PBE} & \makecell{EOM-CCSD \\ in CAM-B3LYP} & \makecell{$\Delta$CAM-\\B3LYP} \\  
\midrule
CH$_2$O + 5H$_2$O  & 1s(C)$\rightarrow$$\infty$& 1 & 295.24 & 0.10 & 0.10 & -0.33 \\
&& 2                             & 295.27 & 0.05 & 0.06 & -0.30 \\
&& 3                             & 295.25 & 0.05 & 0.06 &-0.31 \\
&& 4                             & 295.79 & 0.02 & 0.06 &-0.32 \\[0.3cm]
CH$_2$O + 5H$_2$O &1s(O)$\rightarrow$$\infty$& 1 & 540.98 & 0.11 & 0.12 & -1.39 \\
&& 2                            & 540.95 & 0.04 & 0.06  & -1.40 \\
&& 3                            & 540.95 & 0.05 & 0.06  & -1.40 \\
&& 4                            & 541.48 & 0.01 & 0.06  & -1.46 \\[0.3cm]
HCF + 5H$_2$O  & 1s(C)$\rightarrow$$\infty$ &  & 296.26 & 0.16  & 0.37 & 5.06 \\[0.3cm]
CH$_2$  + 5H$_2$O & 1s(C)$\rightarrow$$\infty$ &  & 293.06 & 0.20 & 0.28 & -0.92 \\
\bottomrule
\end{tabular}}
\label{5-cip}
\end{table}

\textbf{Core ionizations.} Table \ref{5-cip} illustrates that embedded EOM-IP-CCSD performs 
on average somewhat better for core ionization than for valence ionization. The performance 
depends again only weakly on the low-level method used to describe the environment. Embedded 
EOM-IP-CCSD is significantly more accurate than $\Delta$DFT for core ionization, which is in 
contrast to what we saw for valence ionizations. However, is it necessary to consider here that the 
reference values (full fc-CVS-EOM-CCSD) are obtained with the CVS scheme \cite{Vid2019} and 
the $\Delta$DFT values without. For some cases with an environment consisting of a single H$_2$O 
molecule, we found that fc-CVS-EOM-CCSD deviates from EOM-CCSD by up to 0.5 eV and the 
relatively large $\Delta$DFT errors in Table \ref{5-cip} might be partly a consequence of using 
fc-CVS-EOM-CCSD as reference. 

A systematic investigation of this effect was, however, not possible for technical reasons: Since 
core-ionized states are unstable towards Auger decay, they are embedded in the continuum \cite{Jag2017}. 
This leads to convergence problems for the EOM-CCSD equations of core ionized states \cite{Sad2017}; 
for the larger clusters with an environment of 5 H$_2$O molecules, we were unable to achieve 
convergence. With the CVS scheme,\cite{Ced1980} the continuum is projected out from the excitation 
space and the core-ionized states can be obtained without problems. We add here that CVS-EOM-CCSD 
and regular EOM-CCSD both represent approximations because they ignore the resonance character 
of core-ionized states. However, whereas CVS is a well-defined approximation, the approximation in 
regular EOM-CCSD is less obvious and, in fact, uncontrolled because the results depend on how the 
continuum is discretized by the finite basis set. 


\begin{table} \setlength{\tabcolsep}{5pt}
\tbl{Electron attachment energies calculated with full and embedded EOM-EA-CCSD/aug-cc-pVTZ 
and with $\Delta$DFT. Column headings ``full'' and ``trunc.'' refer to calculations using the full and 
a truncated virtual orbital space, respectively. All values in eV.}
{\begin{tabular}{lccccccccccc} \toprule
\multirow{2.5}{*}{System} & \multirow{2.5}{*}{Transition}&\multirow{2.5}{*}{Configuration} & 
\multirow{2.5}{*}{\makecell{full\\EOM-\\CCSD}}& \multicolumn{2}{c}{\makecell{EOM-CCSD \\ in PBE}} & 
\multicolumn{2}{c}{\makecell{EOM-CCSD \\ in CAM-B3LYP}} & \makecell{EOM-CCSD \\ in HF} & 
\makecell{$\Delta$CAM-\\B3LYP}   \\  \cmidrule{5-9}
&&&&full & trunc. &full & trunc. & full &\\ \midrule
HCF + 5H$_2$O &$\infty$$\rightarrow$$\pi$*& 1  &0.34\textsuperscript{a}& -0.60 & -0.21 & 
0.06\textsuperscript{a} & 0.40\textsuperscript{a}  & 0.56\textsuperscript{a} & -0.21 \\
&& 2           & 0.24\textsuperscript{a}& -1.03 & -0.25 & -0.19 & 0.23\textsuperscript{a} & 
0.33\textsuperscript{a} & -0.30\\
&& 3           & 0.27\textsuperscript{a}& -0.18 & -0.29 & -0.15  & 0.34\textsuperscript{a} & 0.47\textsuperscript{a} & -0.93\\
&& 4           &0.13\textsuperscript{a}& -0.96 & -0.46 & -0.50 & 0.21\textsuperscript{a} & 
0.39\textsuperscript{a} & -0.12\\
&& 5           & -0.07 & -1.42 & -1.06 & -0.41 & -0.01 & 0.16 & -1.33 \\[0.3cm]
CH$_2$ + 5H$_2$O & $\infty$$\rightarrow$$\pi$ && 0.36\textsuperscript{a} & -0.64 & 
0.001\textsuperscript{a} & 0.04\textsuperscript{a} & 0.56\textsuperscript{a}  & 
0.58\textsuperscript{a} & 0.03\textsuperscript{a}\\[0.3cm]
SiH$_2$  + 5H$_2$O & $\infty$$\rightarrow$$\pi$&&-0.28& -0.57 & -0.36 & 
-0.25 & -0.17 & 0.05\textsuperscript{a} & 0.30\textsuperscript{a}     \\
\bottomrule
\end{tabular}}
\tabnote{\textsuperscript{a}Positive electron-attachment energies correspond to unbound states. 
The physical significance of these discretized continuum states is limited.}
\label{5-ea}
\end{table}

\textbf{Electron attachment.} Table \ref{5-ea} presents our results for electron attachment energies 
calculated with embedded EOM-EA-CCSD. It can be seen that, in general, the EA variant performs 
significantly worse than the EE and IP variants. In fact, when tackling electron attachment, high 
accuracy is always needed because attachment energies are typically smaller than excitation or 
ionization energies. Thus, even a small error can have strong repercussions on the results. Table 
\ref{5-ea} shows that several states that are unbound according to full EOM-EA-CCSD 
become artificially bound with embedded EOM-EA-CCSD.

This problem is already present at the self-consistent field level: We observe in many embedding 
calculations virtual orbitals with negative energies emerging that are not present in regular self-consistent 
field calculations. The inclusion of very diffuse functions in the basis set, as is required for Rydberg 
states and resonances, appears to trigger the emergence of these orbitals. They cannot be easily 
recognized by their shape or in some other way but electron attachment to them and excitations 
into them are not physically meaningful because no corresponding orbitals exist in the self-consistent 
field calculations for the full system. In our EOM-EE-CCSD calculations for Rydberg states, excitations 
involving these orbitals were easy to discern by the unphysical shape of the corresponding NTOs but 
this may not be the case if the environment is more complex and consists not only of water molecules.

In general, electron attachment energies depend much more strongly on the density 
functional used to describe the environment than other transition energies. One might expect that 
long-range corrected functionals improve the description \cite{Vyd2007,Jen2010} but we found that 
the problems largely persist with CAM-B3LYP. Also here, virtual orbitals with an artificial negative 
energy are obtained. A significant improvement is obtained when the low-level fragment is described 
at the HF level, but these calculations suffer in a few cases from the opposite problem, that is, a state 
that is bound according to full EOM-EA-CCSD becomes unbound.

Table \ref{5-ea} also shows that the results of embedded EOM-EA-CCSD calculations depend 
significantly on the truncation of the virtual orbital space. This is in contrast to all other types of 
states investigated in this work and will be discussed in more detail in Sec. \ref{sec:trunc}. Contrary 
to the intuition, more accurate attachment energies are in most cases obtained with 
a truncated virtual orbital space. This is probably due to error cancellation and we conclude that care 
should be applied when using embedded EOM-EA-CCSD. 


\subsection{Truncation of the virtual orbital space} \label{sec:trunc}
In order to carry out embedded EOM-CCSD calculations with large environments, it is critical to 
truncate the virtual orbital space. As outlined in Section \ref{sec:comp}, we use here concentric 
localization \cite{Cla2019} to this end. To test the validity of this procedure for different types of 
target states, we compare energies obtained with truncated and untruncated virtual orbital spaces. 
Our numerical results can be found in the SI and are discussed in the following.

We observe that the impact of the truncation is smallest for valence and core ionization energies. 
Here, the errors caused by concentric localization are consistently smaller than 10$^{-5}$ a.u. This 
is not a surprise because the virtual orbitals are not needed for a zeroth-order description of ionized 
states and enter only at higher orders. Valence excitation energies are not heavily affected by the 
truncation either; we observe a maximum deviation of 0.01 eV among our test cases for the transition 
in HCF. 

The excitation energy of CH$_3$OH is an exception as it exhibits truncation errors of 0.1 eV and 
0.05 eV for EOM-CCSD in PBE and CAM-B3LYP, respectively. We relate this to the partial Rydberg 
character of this state because we observe in general larger truncation errors when the electron is 
excited to more spatially extended orbitals. While the performance of concentric localization is overall 
still satisfactory for Rydberg states with deviations of 0.01--0.02 eV in many cases, there are some 
cases with significantly larger deviations. For the 3$p_y$ state of formaldehyde, concentric localization 
leads to a difference of 0.08 eV in the excitation energy. Moreover, there are no clear trends: Due to 
different orientations and a varying degree of delocalization over the environment molecules, the 
truncation process can yield different results for each state in a Rydberg series. 

We also notice that excitation energies are less sensitive towards truncation of the virtual orbital 
space when CAM-B3LYP is employed as low-level method for the environment as compared to 
PBE. The same trend is observed for the intramolecular charge-transfer state in HCl: the difference 
in energies with and without the application of concentric virtual localization is 0.05, 0.04, and 0.01 
eV for embedding in PBE, BLYP and CAM-B3LYP, respectively. 

Most sensitive towards truncation of the virtual orbital space are electron attachment energies. 
Here, the error due to concentric localization is of the same order of magnitude as the one due 
to embedding itself and the results do not change significantly until the full virtual orbital space is 
recovered. This may be related to the fact that already a zeroth-order description of electron-attached 
states involves the virtual orbitals. As alluded to in the previous section (see Table \ref{5-ea}), 
concentric localization improves the agreement with full EOM-EA-CCSD in many 
cases, likely due to error cancellation. However, the performance is still inferior 
to the other embedded EOM-CCSD variants and further investigations are needed to exploit the 
cancellation of errors in a useful way.

As a final remark, we note that EOM-CCSD calculations with a truncated virtual space appear to be 
more likely to converge to a solution that is dominated by an excitation involving an artificial virtual 
orbital than calculations with the full virtual space. However, concentric localization is not the origin 
of this problem as virtual orbitals with artificial negative energies are also present without truncation. 

\subsection{Natural transition orbitals and Dyson orbitals}

\begin{figure}[h!] \centering
\includegraphics[scale=0.10, trim = 510 50 510 50, clip]{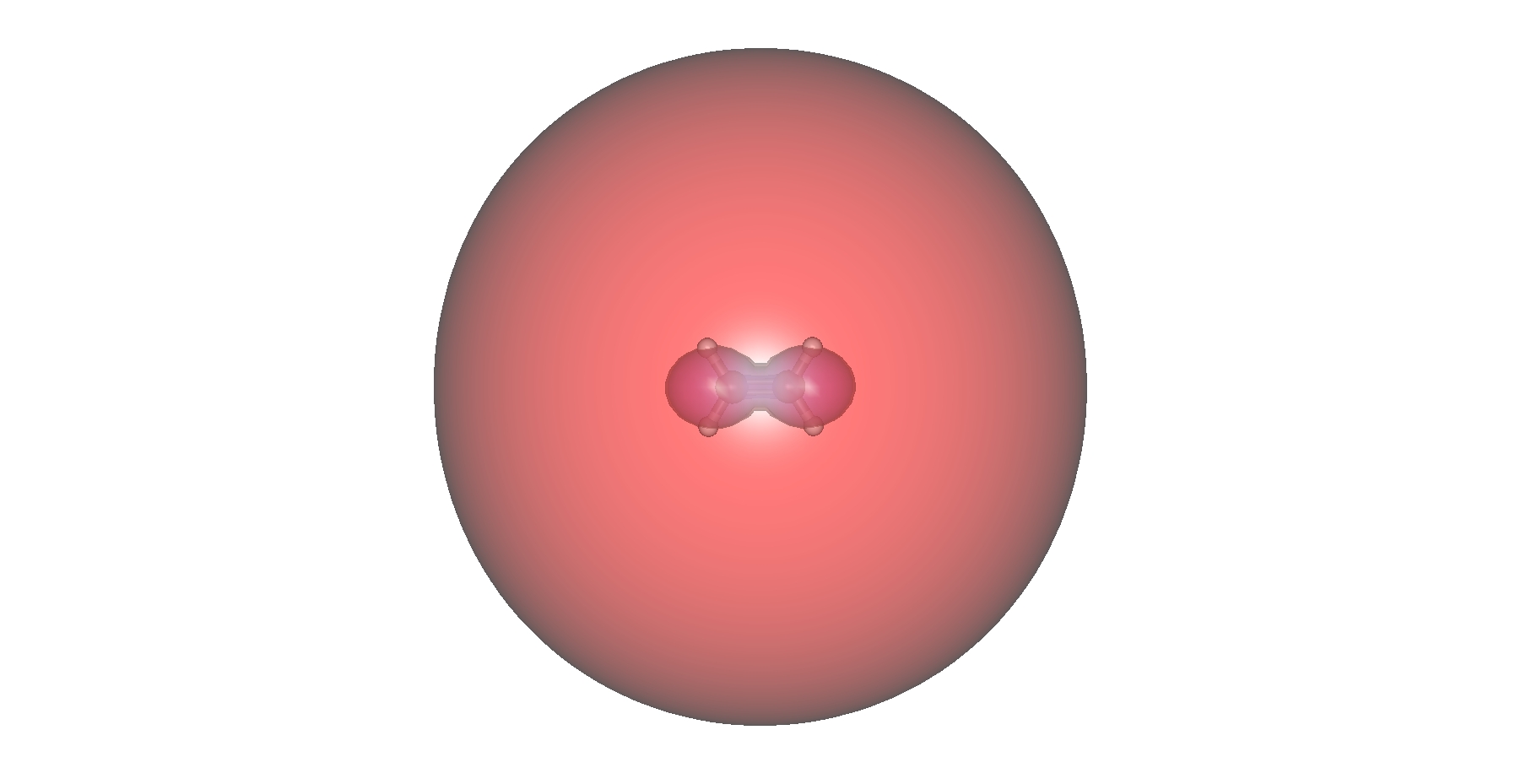} \hspace{0.75cm}
\includegraphics[scale=0.10, trim = 540 0 540 0, clip]{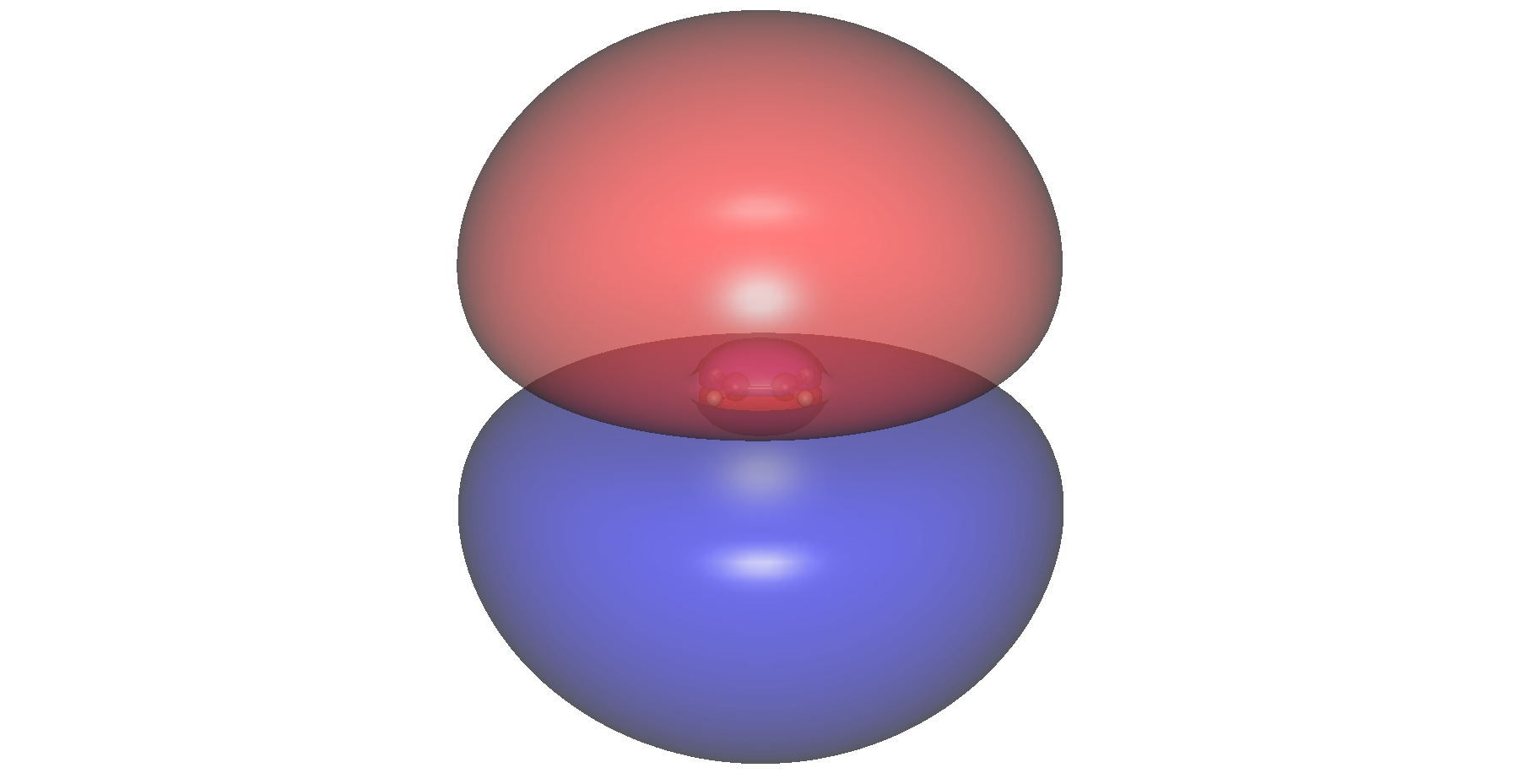} \hspace{0.75cm}
\includegraphics[scale=0.10, trim = 540 20 540 20, clip]{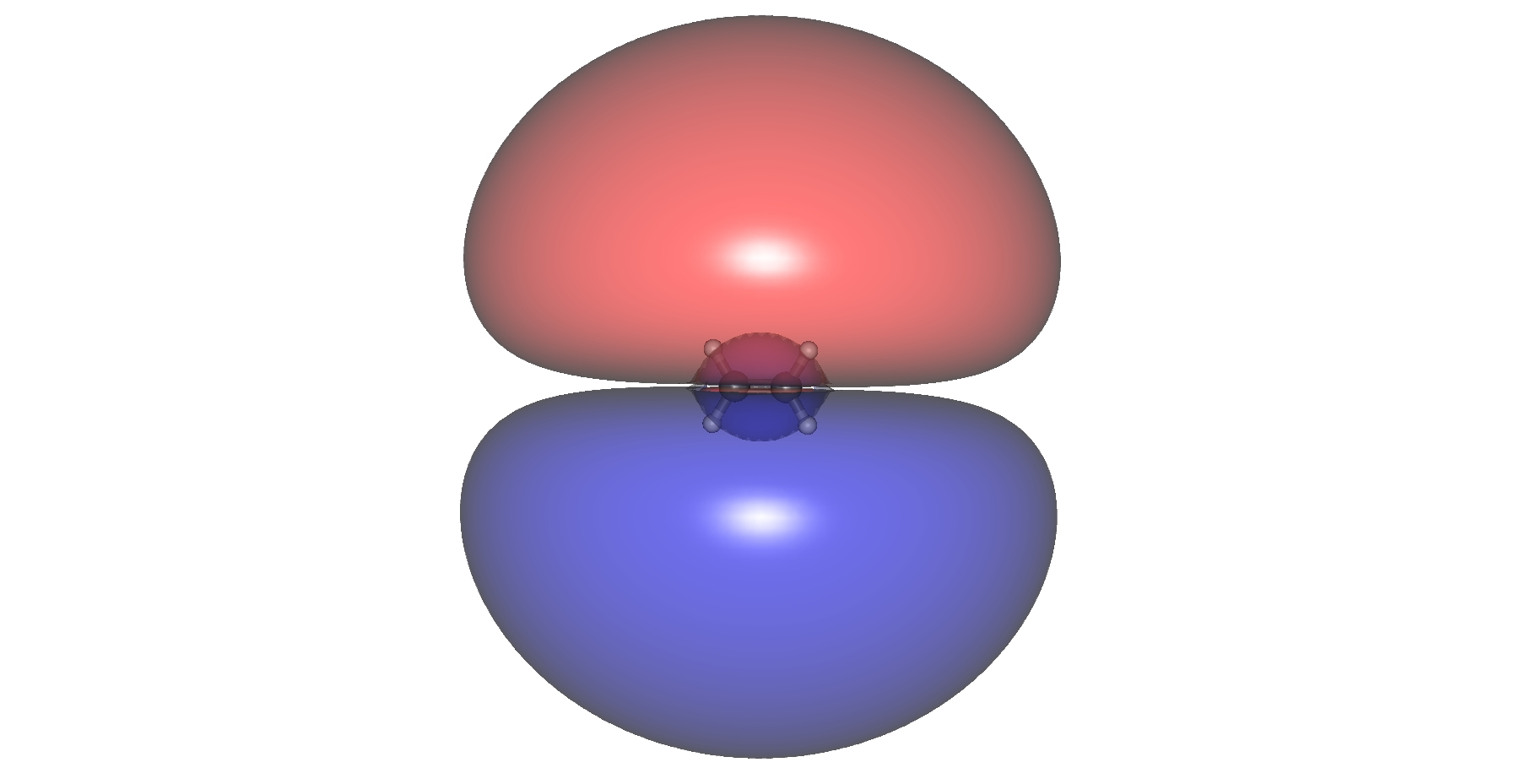} \hspace{0.75cm}
\includegraphics[scale=0.09, trim = 360 20 360 20, clip]{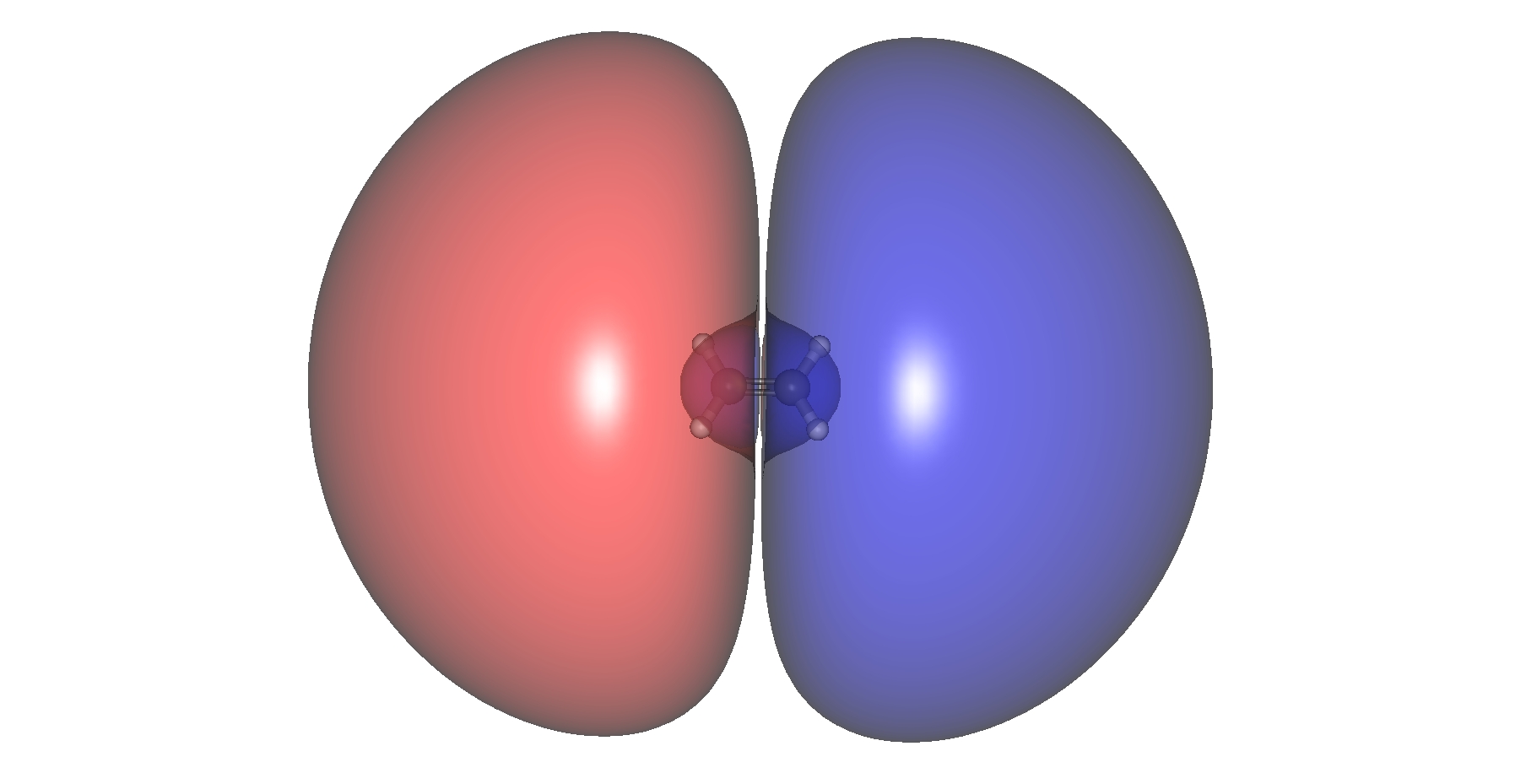}  \\[1.0cm]

\includegraphics[scale=0.12, trim = 550 70 550 70, clip]{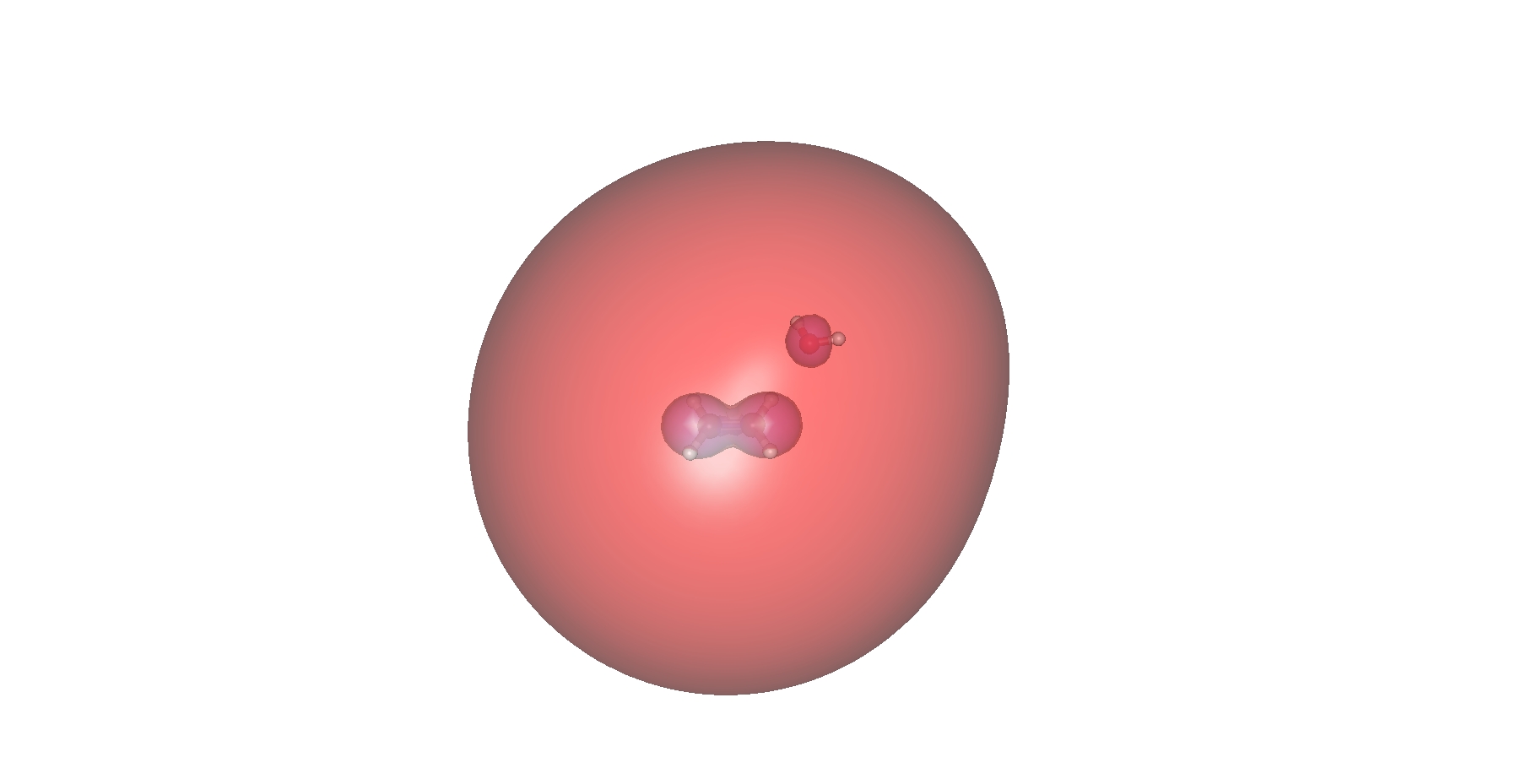} \hspace{0.2cm}
\includegraphics[scale=0.11, trim = 500 20 500 20, clip]{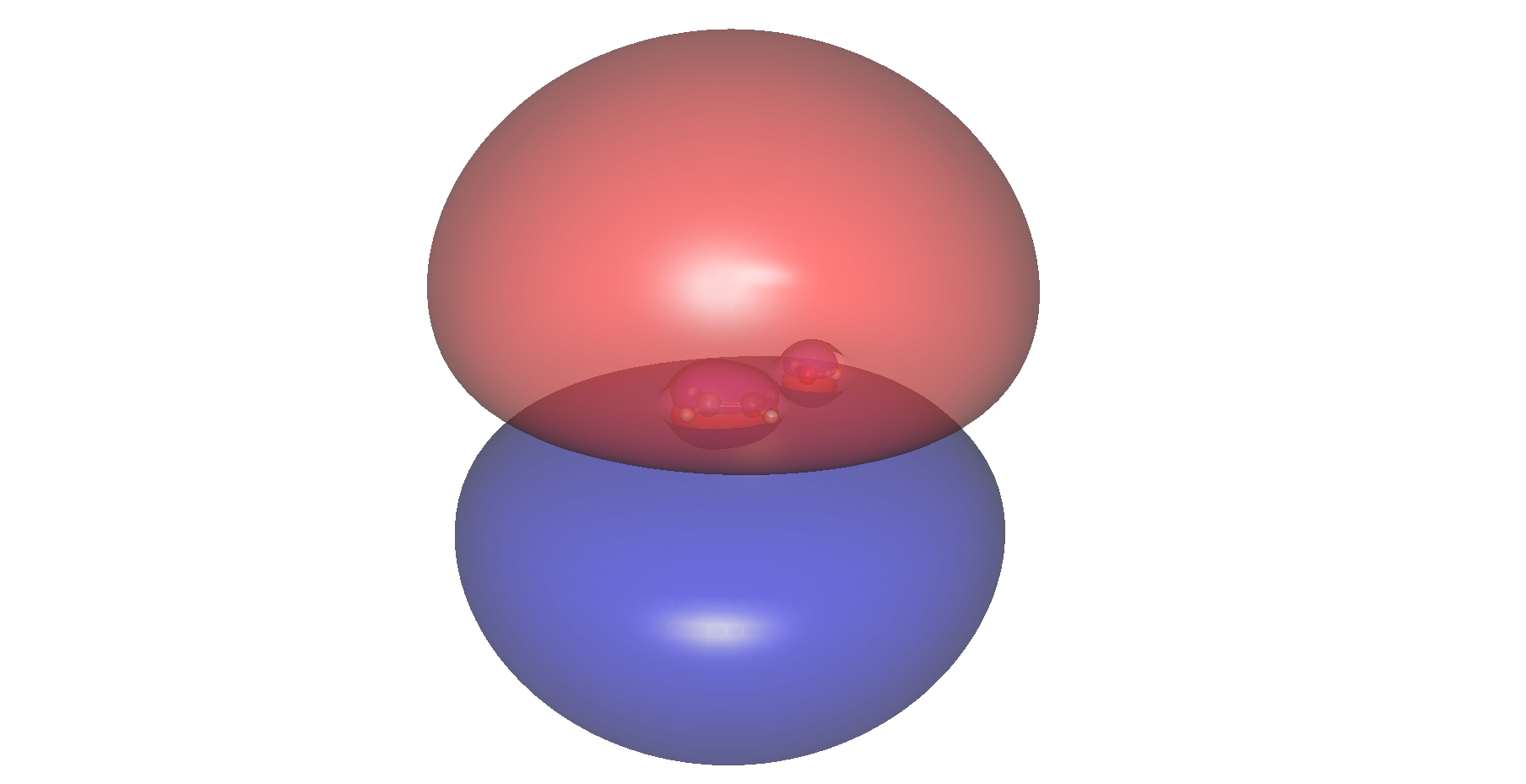} \hspace{0.2cm}
\includegraphics[scale=0.11, trim = 480 30 480 30, clip]{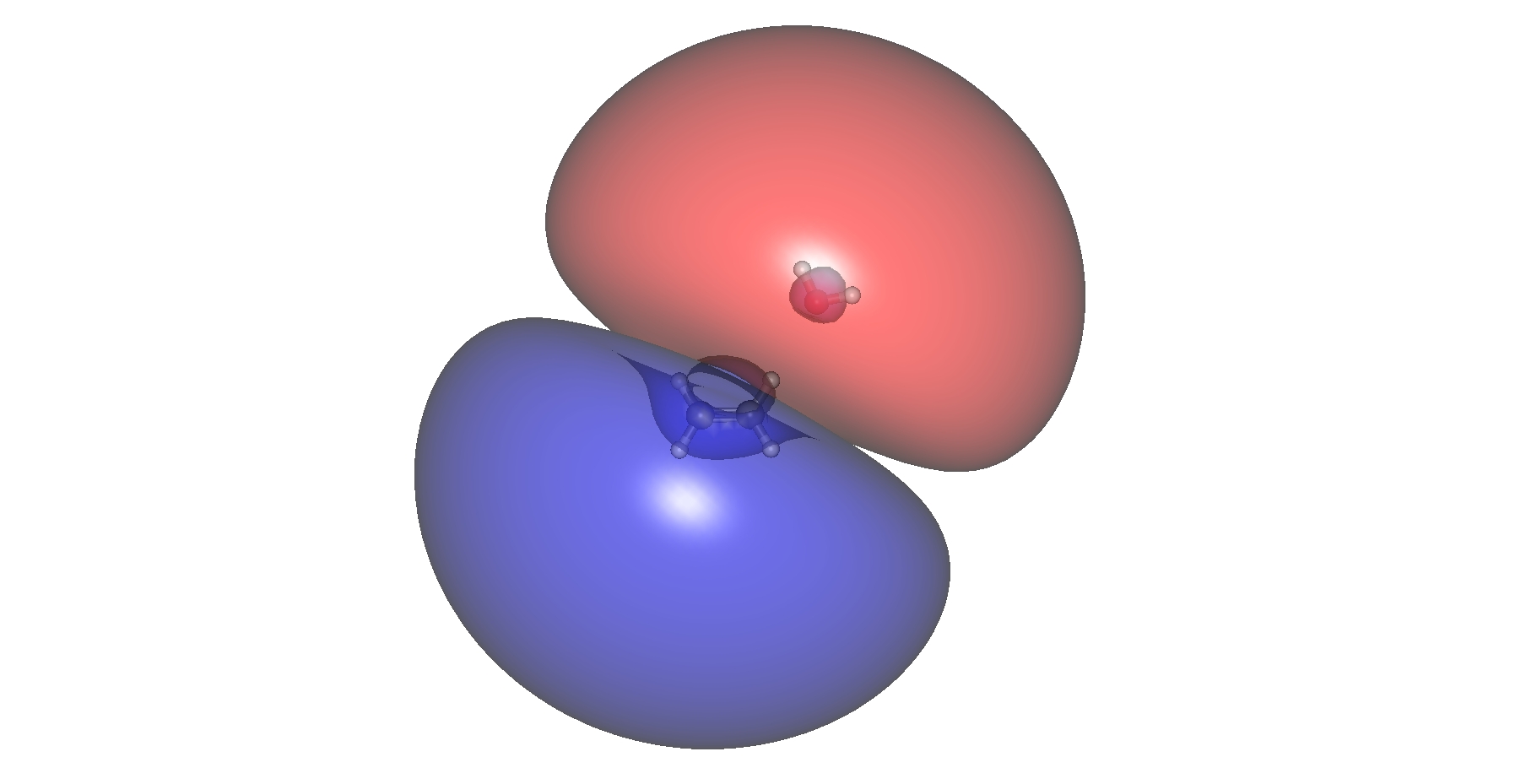} \hspace{0.2cm}
\includegraphics[scale=0.10, trim = 380 20 420 20, clip]{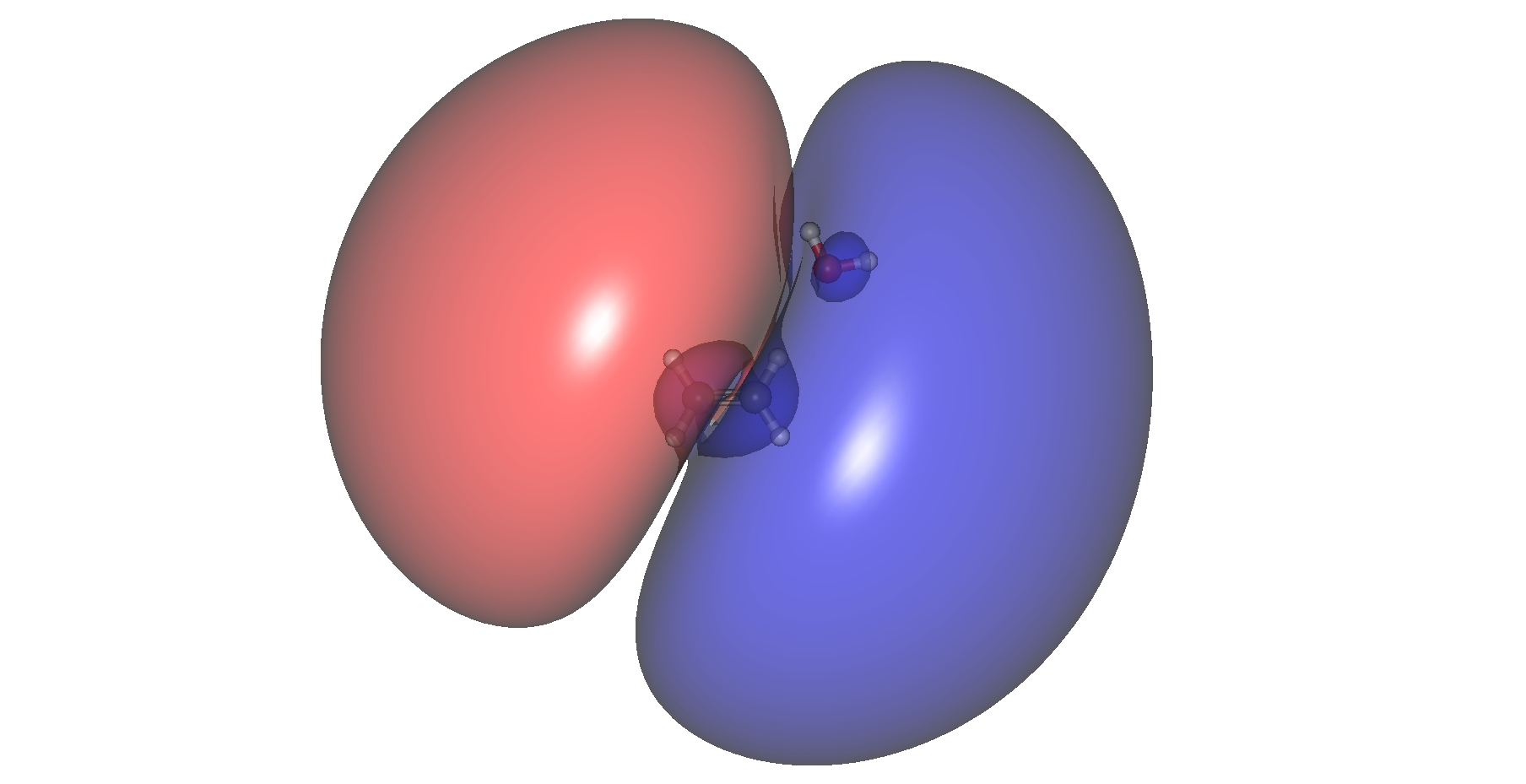}

\caption{Upper row: Particle NTOs for excitations to the 3s, 3p$_x$, 3p$_y$, and 
3p$_z$ Rydberg orbitals of C$_2$H$_4$ computed with EOM-EE-CCSD/aug-cc-pVTZ+2s2p. 
Lower row: Corresponding particle NTOs for C$_2$H$_4$ + H$_2$O computed with EOM-EE-CCSD 
embedded in PBE0/aug-cc-pVTZ+2s2p. All NTOs have been plotted at an isovalue of 0.001.} 
\label{fig:nto}
\end{figure}

Figure \ref{fig:nto} shows the particle NTOs for several Rydberg states of isolated ethylene and 
of ethylene in the presence of a water molecule located nearby in the molecular 
plane ($yz$-plane). It can be seen that the environment participates in the transition to a varying 
degree: none of the NTOs is unaffected by the presence of the water molecule but 
the NTOs of the 3p$_y$ and the 3p$_z$ state are more significantly distorted than those of the 3s 
state and the 3p$_x$ state. 

To quantify the similarity between NTOs from embedded and full EOM-EE-CCSD calculations, 
we compare in Table \ref{eigenval} the largest singular values of the transition density matrices. 
Overall, excellent agreement between the two approaches is observed. Furthermore, we notice 
that when the difference in the transition energy increases, so does the difference between the 
singular values. The largest deviations are observed for the 3p$_y$ and the 3p$_z$ 
state in agreement with Fig. \ref{fig:nto}.

\begin{table}[h!]
\tbl{Energies $E$ and largest singular values $\sigma$ of the transition density matrix computed 
with full EOM-EE-CCSD for the Rydberg series of C$_2$H$_4$+H$_2$O and deviations $\Delta E$ 
and $\Delta \sigma$ for EOM-EE-CCSD embedded in PBE0. All values are computed using the 
aug-cc-pVTZ+2s2p basis set.}
{\begin{tabular}{lcccc} \toprule
 & \multicolumn{2}{c}{Full EOM-EE-CCSD} & \multicolumn{2}{c}{EOM-EE-CCSD in PBE0} \\
Transition & $E$ / eV &  $\Delta E$ / eV &  $\sigma$ / a.u. & $\Delta \sigma$ / a.u. \\
\midrule
$\pi\rightarrow$3s     & 7.15 & 0.07 & 0.87 & 0.006 \\
$\pi\rightarrow$3p$_y$ & 7.72 & 0.16 & 0.86 & 0.013 \\
$\pi\rightarrow$3p$_x$ & 7.84 & 0.11 & 0.86 & 0.010 \\
$\pi\rightarrow$3p$_z$ & 8.10 & 0.01 & 0.87 & 0.004 \\
\bottomrule
\end{tabular}} \label{eigenval}
\end{table}

\begin{figure}[h!] \centering
\includegraphics[scale=0.14, trim = 320 40 320 80, clip]{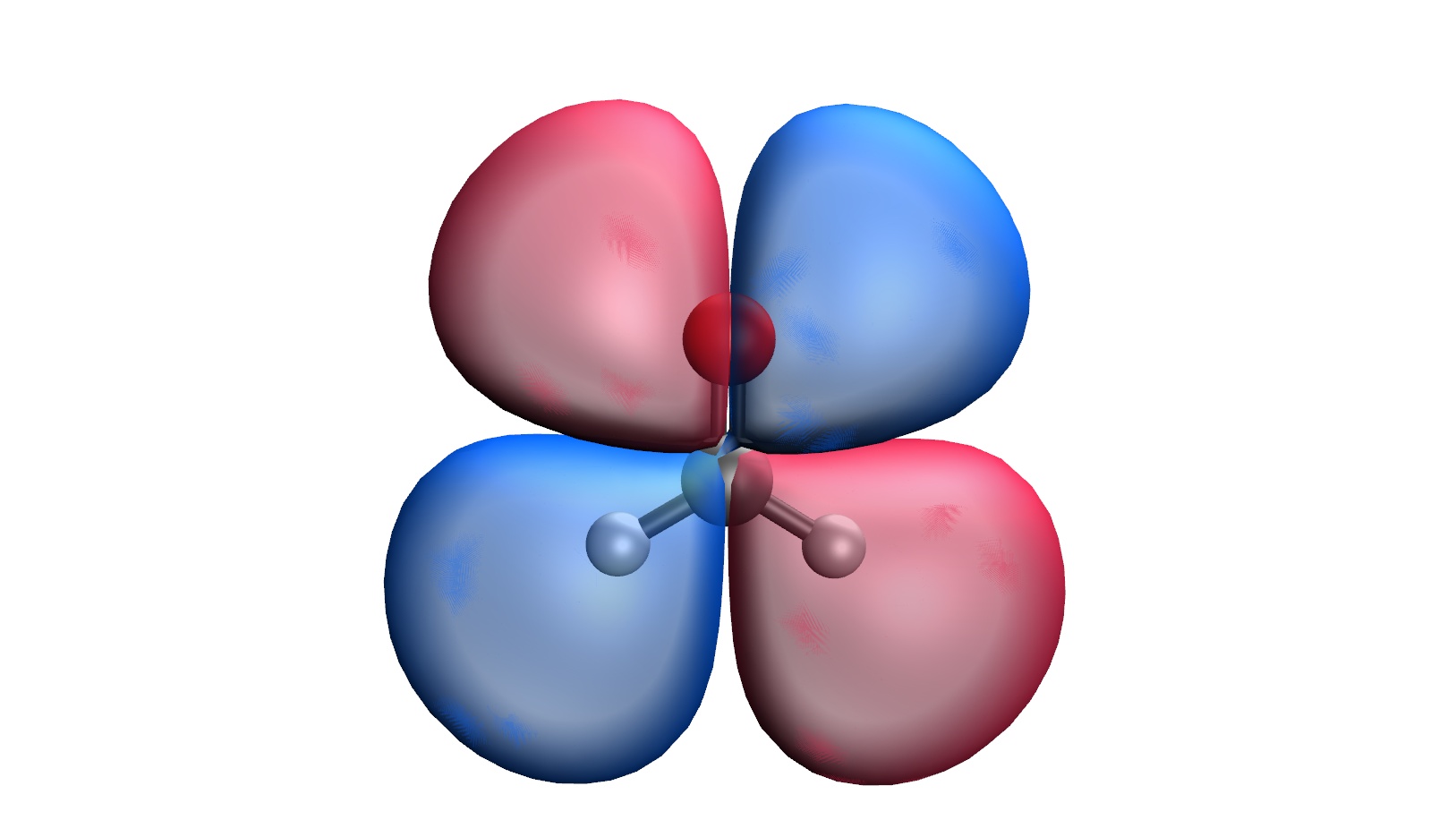} \hspace{2cm}
\includegraphics[scale=0.15, trim = 250 60 250 60, clip]{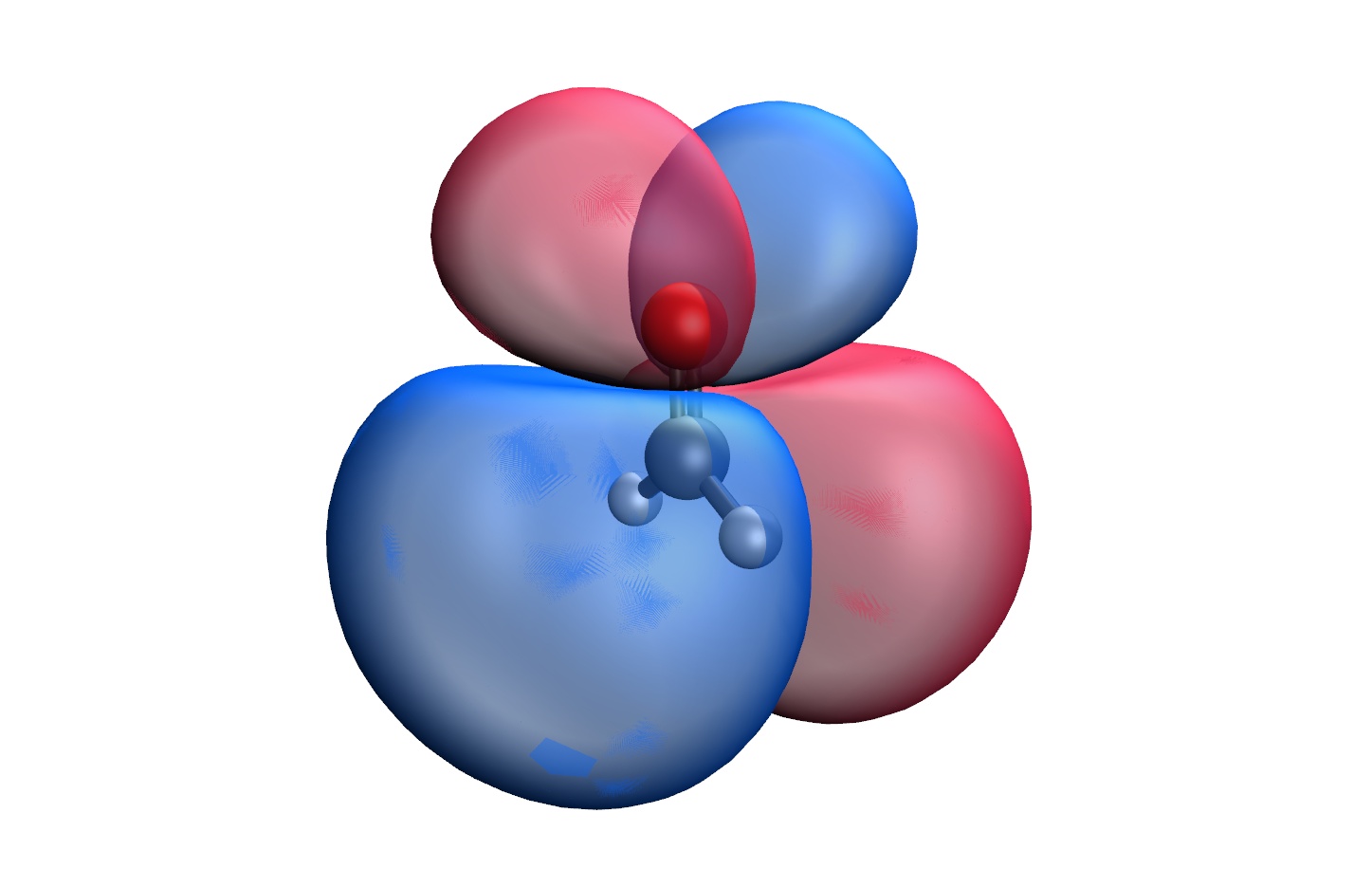} \hspace{2cm}
\includegraphics[scale=0.08, trim = 400 0 400 100, clip]{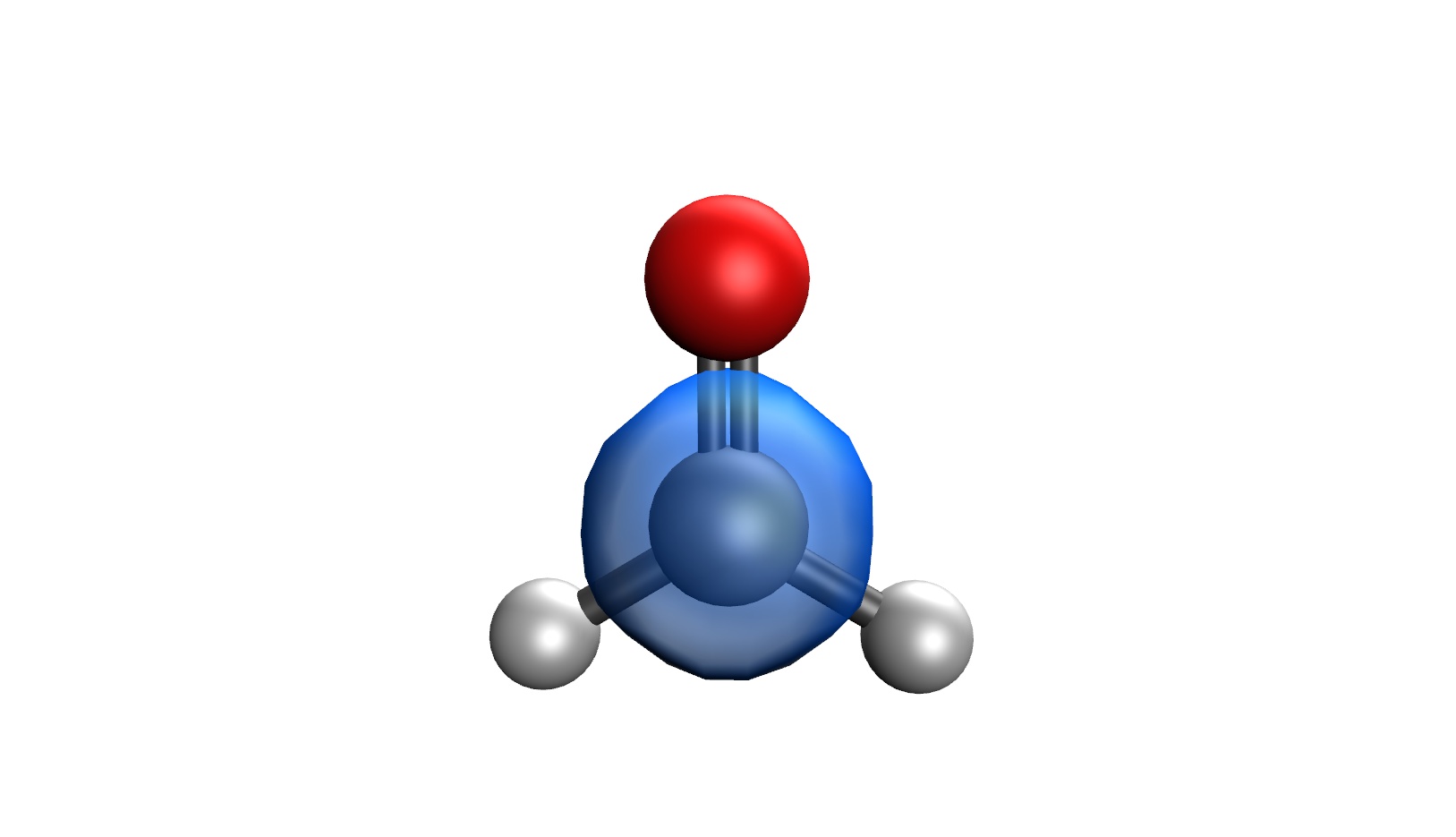} \\[0.8cm]

\includegraphics[scale=0.25, trim = 100 20 100 20, clip]{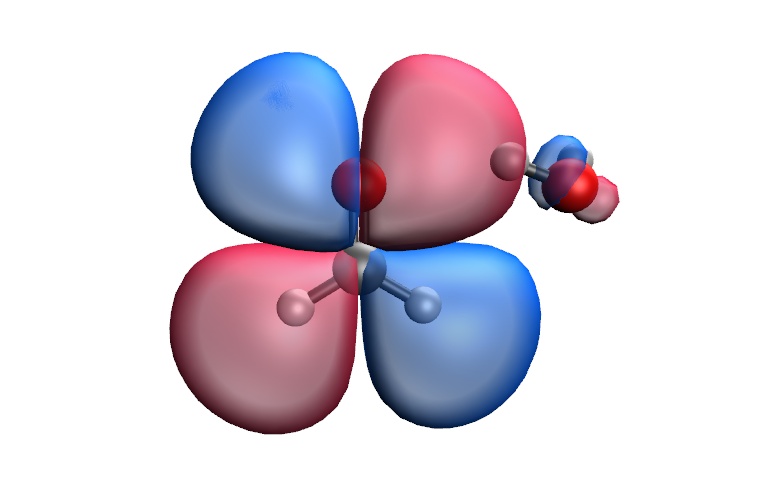} \hspace{2cm}
\includegraphics[scale=0.20, trim = 200 40 200 60, clip]{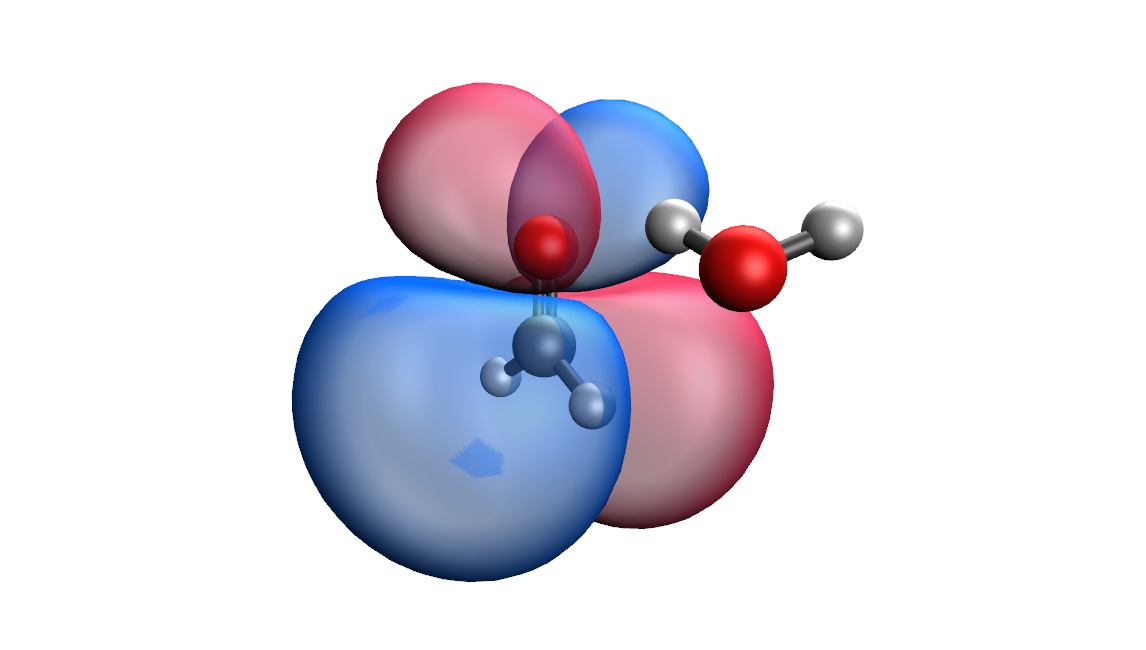} \hspace{2cm}
\includegraphics[scale=0.19, trim = 100 0 120 80, clip]{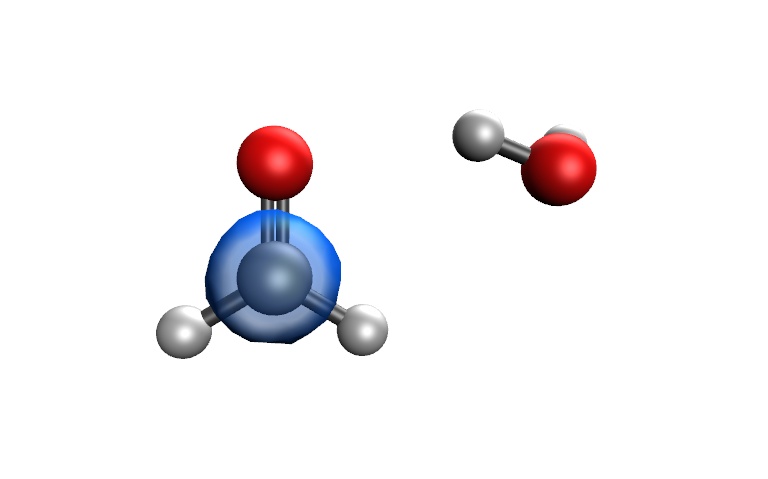} 

\caption{Upper row: Dyson orbitals for valence ionization of the ground state, valence 
ionization of the lowest excited state, and 1s(C) core ionization of CH$_2$O computed 
with EOM-IP-CCSD/aug-cc-pVTZ. Lower row: Corresponding Dyson orbitals for CH$_2$O 
+ H$_2$O computed with EOM-IP-CCSD embedded in PBE0/aug-cc-pVTZ. The 
frozen-core-CVS scheme has been employed for the core-ionized state. Dyson 
orbitals describing valence ionization have been plotted at an isovalue of 0.005, those 
describing core ionization at an isovalue of 0.007.} 
\label{fig:dy}
\end{figure}

As a further example, we present Dyson orbitals for valence and core ionization of CH$_2$O and 
CH$_2$O + H$_2$O in Figure \ref{fig:dy}. The comparison to Figure \ref{fig:nto} shows that Dyson 
orbitals are, in general, less affected by the environment than particle NTOs. Only for the Dyson 
orbital that describes valence ionization of the ground state, some participation of the environment 
can be seen in the plot. On the contrary, no difference is visible for the Dyson orbitals that describe 
valence ionization of the first excited state and core ionization of the ground state. Consistent with 
this, the ionization energy corresponding to the first Dyson orbital changes by almost 0.4 eV in the 
presence of the H$_2$O molecule while the other two ionization energies change by only 0.26 eV 
and 0.22 eV. 


\section{Results for electronic resonances} \label{sec:results2}

\subsection{Localized electron attachment}
We test the performance of projection-based embedding combined with CAP-EOM-EA-CCSD first 
on the temporary anions CO$^-$ and CH$_2$O$^-$ solvated by one water molecule. In these 
systems, the extra electron occupies the $\pi$* orbital of the C--O bond of CO or CH$_2$O, that is, 
the excess electron density is localized.

From the data in Table \ref{cap-loc}, we observe that CAP-EOM-EA-CCSD embedded in PBE 
reproduces full CAP-EOM-EA-CCSD results with high accuracy ($10^{-2}$ eV) for positions and 
widths. The results are also largely independent of the CAP shape. This good agreement of energies 
and decay widths obtained with embedding justifies in retrospect our approach to apply the CAP only 
to the EOM-CCSD part. However, the deviation between embedded CAP-EOM-EA-CCSD and full 
EOM-EA-CCSD increases when the distance between the transient anion and water decreases. 
This is in line with our results obtained for bound anions microsolvated by one molecule of water. 

\begin{table}[h!] 
\tbl{Resonance positions and widths of CH$_2$O$^-$ and CO$^-$ microsolvated by one H$_2$O 
molecule computed with embedded and full CAP-EOM-EA-CCSD/aug-cc-pVTZ+3s3p at different 
intermolecular distances. Box CAP onsets and optimal $\eta$ values are reported in the SI. All 
values in eV.}
{\begin{tabular}{lcccccc} \toprule
\multirow{2.5}{*}{System}& Intermolecular & \multicolumn{2}{c}{CAP-EOM-CCSD} & 
\multicolumn{2}{c}{CAP-EOM-CCSD in PBE} \\
\cmidrule{3-6}
 & Distance/\si{\angstrom} & $E_R$ & $\Gamma$ & $E_R$ & $\Gamma$ \\
\midrule
box CAP \\
\midrule
CH$_2$O$^-$ &&1.07 & 0.38 & & \\
CH$_2$O$^-$ + H$_2$O & 2.2 & 0.88 & 0.20 & 0.91 & 0.22 \\[0.3cm]
CO$^-$ & &2.07 & 0.60 \\
CO$^-$ + H$_2$O & 2.0 & 1.68 & 0.47 & 1.72 & 0.55 \\
CO$^-$ + H$_2$O & 2.5 & 1.82 & 0.56 & 1.85 & 0.58 \\
CO$^-$ + H$_2$O & 3.0 & 1.91 & 0.55 & 1.92 & 0.55 \\
CO$^-$ + H$_2$O & 4.0 & 1.99 & 0.53 & 1.99 & 0.54 \\
\midrule
\multicolumn{3}{l}{Voronoi CAP (cutoff: 4 \si{\angstrom})} \\
\midrule
CO$^-$ + H$_2$O & 2.0 & 1.67 & 0.56 & 1.72 & 0.52 \\
CO$^-$ + H$_2$O & 2.5 & 1.82 & 0.56 & 1.83 & 0.55 \\
CO$^-$ + H$_2$O & 3.0 & 1.90 & 0.57 & 1.90 & 0.60 \\
CO$^-$ + H$_2$O & 4.0 & 1.99 & 0.60 & 1.98 & 0.62 \\
\bottomrule
\end{tabular}} 
\label{cap-loc} 
\end{table}

\subsection{Delocalized electron attachment}
When the excess electron density is delocalized over the environment, embedded CAP-EOM-EA-CCSD 
does not perform equally well. Here, we report results for the temporary anions (CO)$_2^-$ and 
(N$_2$)$_2^-$. Four resonance states can be identified in these dimers; they arise from the combination 
of the $\pi$* orbitals in CO or N$_2$, respectively. We note that (CO)$_2^-$ has been studied before in 
Ref. \citenum{Gayv2020}, here we use the same geometry. 

To test the applicability of projection-based embedding, only one of the monomers is included in 
the CAP-EOM-EA-CCSD calculation, while the other one constitutes the environment. We do not 
truncate the virtual orbital space in these calculations, so the virtual orbitals of both monomers 
take part in the CAP-EOM-CCSD treatment. The density functionals PBE and CAM-B3LYP are 
assessed for both dimers, for (CO)$_2^-$ we investigate in addition HF, $\omega$B97X-D, and 
LC-$\omega$PBE08.

In none of these cases are the results satisfactory. With PBE as low-level method, artificial bound 
states appear, i.e., states with a negative attachment energy, thus making it inappropriate for the 
description of such systems. This is reminiscent of the problems of EOM-EA-CCSD embedded in 
PBE that we observed for bound anions (see Sec. \ref{sec:results1}). The problem with artificial 
bound states is not present with the other density functionals or HF theory, which confirms our 
previous finding that longe-range corrected functionals improve the performance of embedded 
EOM-EA-CCSD. However, large deviations from full CAP-EOM-EA-CCSD still persist as demonstrated 
by Table \ref{cap-deloc}. Therefore, we conclude that for a correct description of temporary anions 
with a delocalized excess electron density, all parts of the system have to be treated on an equal 
footing. 

\begin{table}[h!] \setlength{\tabcolsep}{2.5pt}
\tbl{Resonance positions and widths of (CO)$_2^-$ and (N$_2$)$_2^-$ computed with embedded 
and full CAP-EOM-EA-CCSD/aug-cc-pVTZ+3s3p. Box CAP onsets and optimal $\eta$ values are 
reported in the SI. All values in eV.}
{\begin{tabular}{ccccccccccccc} \toprule
 & \multicolumn{2}{c}{CAP-EOM-CCSD} & \multicolumn{2}{c}{CAP-EOM-CCSD} & 
 \multicolumn{2}{c}{CAP-EOM-CCSD} & \multicolumn{2}{c}{CAP-EOM-CCSD} & 
 \multicolumn{2}{c}{CAP-EOM-CCSD} \\
 &  &  & \multicolumn{2}{c}{in CAM-B3LYP} & \multicolumn{2}{c}{in $\omega$B97X-D} & 
\multicolumn{2}{c}{in LC-$\omega$PBE08} & \multicolumn{2}{c}{in HF} \\ \midrule
State & $E_R$ & $\Gamma$ & $E_R$ & $\Gamma$ & $E_R$ & $\Gamma$ & $E_R$ & 
$\Gamma$ & $E_R$ & $\Gamma$ \\ \midrule
 & \multicolumn{4}{c}{(CO)$_2^-$} \\[0.1cm] 
A$_u$ & 1.40 & 0.92 & 0.94 & 0.23 & 0.82 & 0.28 & 1.56 & 0.17 & 1.92 & 0.60 \\
A$_g$ & 1.72 & 0.23 & 1.07 & 0.19 & 0.96 & 0.10 & 1.57 & 0.41 & 2.05 & 0.27\\
B$_g$ & 2.31 & 0.29 & 2.50 & 0.58 & 2.08 & 0.52 & 2.21 & 0.32 & ---\textsuperscript{a} & 
---\textsuperscript{a} \\
B$_u$ & 2.91 & 0.51& 2.78 & 0.70 & 2.45 & 0.49 & 2.68 & 0.53& ---\textsuperscript{a} & 
---\textsuperscript{a} \\ \midrule
 & \multicolumn{4}{c}{(N$_2$)$_2^-$} \\[0.1cm] 
1A$^{\prime}$ & 2.28 & 0.44 & 1.13 & 0.08 &  &  \\
1A$^{\prime\prime}$& 2.32 & 0.61 & 1.25 & 0.15 &  &  \\
2A$^{\prime\prime}$& 2.62 & 0.24 & 2.45 & 0.59 &  &  \\
2A$^{\prime}$ & 2.93 & 0.43 & 2.65 & 0.40 &  &  \\
\bottomrule
\end{tabular}}
\tabnote{\textsuperscript{a}These resonances could not be observed.}
\label{cap-deloc}
\end{table}


\section{Conclusions} \label{sec:conc}
We have investigated the performance of projection-based embedding by means of a comprehensive 
benchmark set of various types of electronic states. Different flavors of embedded EOM-CCSD were 
applied to microsolvated small organic molecules and the results were compared to full EOM-CCSD. We 
found that embedded EOM-CCSD, in general, works best if the distance between the high-level fragment 
and the environment is large; we also observed a mild systematic deterioration if the environment comprises 
five instead of one single water molecule. 

Most interesting, however, is the comparison between different types of target states: in agreement with 
previous investigations \cite{Ben2017,Wen2020}, we found that embedded EOM-CCSD in general works 
very well for valence excitations. Our investigations show that the same can also be said about valence 
and core ionizations, intramolecular charge transfer, as well as Rydberg excitations, although the use of 
range separated density functionals for the environment proves to be crucial for the latter. Moreover, all 
these types of transitions are fairly robust towards truncation of the virtual orbital space so that the investigation 
of more extended systems, where the environment is larger than in our current work, is straightforward. 
We note that TD-DFT and $\Delta$DFT are competitive alternatives for valence excitations and ionizations. 
The clearest improvement over a simple DFT approach is thus obtained for Rydberg excitations and 
core ionizations according to our results. A further important advantage of embedded EOM-CCSD over 
$\Delta$DFT approaches is the possibility to compute not only transition energies but also further quantities 
such as Dyson orbitals, which we have introduced in this work alongside natural transition orbitals. 

In contrast to ionization and excitation, electron attachment proves to be more challenging to describe 
with embedded EOM-CCSD. The appearance of virtual orbitals with unphysical negative energy 
eigenvalues can in many cases not be avoided, the truncation of the virtual orbital 
space changes attachment energies significantly and, while some results benefit from error cancellation, 
the overall performance of the method is not satisfying. However, the extension from bound to 
temporary anions by means of complex absorbing potentials is straightforward and delivers good 
results for energies and decay widths as long as the excess electron density remains localized. 
Applications to temporary anions are, however, subject to the same limitation as those to bound 
anions: the virtual orbital space cannot be easily truncated. Some refinement of the embedding 
procedure thus appears to be essential to obtain a viable approach for electron attachment.

\section*{Associated content}
The supporting information available for this work contains all molecular 
structures, relevant energies and technical parameters of the complex absorbing potentials calculations.

\section*{Acknowledgments}
This article is dedicated to Professor John Stanton on the occasion of his 60\textsuperscript{th} birthday. 
T.C.J. would like to thank him for many fruitful discussions and valuable support on different occasions. 
We thank Dr. Simon Bennie, Professor Basile Curchod, and Professor Fred Manby for sharing some 
of the raw data used for their work on embedded EOM-CCSD and Professor Anna Krylov 
for valuable feedback on the manuscript. This work has been funded by the European Research Council 
(ERC) under the European Union's Horizon 2020 research and innovation program (Grant Agreement 
No. 851766). The computational resources and services used in this work were partly provided by the 
VSC (Flemish Supercomputer Center), funded by the Research Foundation - Flanders (FWO) and the 
Flemish Government.

\pagebreak 

\end{document}